\documentclass[prx,twocolumn,superscriptaddress]{revtex4}
\usepackage{amsmath}
\usepackage{amssymb}
\usepackage{amsthm}
\usepackage{amsfonts}
\usepackage{listings}
\usepackage{enumerate}
\usepackage{latexsym}
\usepackage{color}
\usepackage{bm}
\usepackage{hyperref}
\hypersetup{
 pdfnewwindow=true, colorlinks=true,
 linkcolor=blue, anchorcolor=blue,
 citecolor=blue, filecolor=blue,
 menucolor=blue, urlcolor=blue}

\newcommand{\beginsupplement}{%
        \setcounter{table}{0}
        \renewcommand{\thetable}{S\arabic{table}}%
        \setcounter{figure}{0}
        \renewcommand{\thefigure}{S\arabic{figure}}%
     }

\usepackage{psfrag}
\usepackage{color}

\usepackage{bm}
\usepackage{graphicx}
\usepackage{subfigure}

\RequirePackage[normalem]{ulem} 
\RequirePackage{color}\definecolor{RED}{rgb}{1,0,0}\definecolor{BLUE}{rgb}{0,0,1} 

\newcommand{\bk}{{\bf k}}
\newcommand{\bg}{{\bf g}}
\newcommand{\bt}{{\bf t}}

\def\ie{{\it i.e.},\ }
\def\eg{{\it e.g.}\ }

\input{epsf}

\begin{document}

\tolerance 10000

\newcommand{\vk}{{\bf k}}

\draft

\title{Symmetry-Enforced Weyl Phonons}

\author{Qing-Bo Liu}
\affiliation{School of Physics and Wuhan National High Magnetic field center,
Huazhong University of Science and Technology, Wuhan 430074, People's Republic of China.}

\author{Yuting Qian}
\affiliation{Beijing National Laboratory for Condensed Matter Physics,
and Institute of Physics, Chinese Academy of Sciences, Beijing 100190, China}
\affiliation{University of Chinese Academy of Sciences, Beijing 100049, China}

\author{Hua-Hua Fu}
\email{hhfu@hust.edu.cn}
\affiliation{School of Physics and Wuhan National High Magnetic field center,
Huazhong University of Science and Technology, Wuhan 430074, People's Republic of China.}

\author{Zhijun Wang}%
\email{wzj@iphy.ac.cn}
\affiliation{Beijing National Laboratory for Condensed Matter Physics,
and Institute of Physics, Chinese Academy of Sciences, Beijing 100190, China}
\affiliation{University of Chinese Academy of Sciences, Beijing 100049, China}


\begin{abstract}
In spinful electronic systems, time-reversal symmetry makes that all Kramers pairs at the time-reversal-invariant momenta are Weyl points (WPs) in chiral crystals. Here, we find that such symmetry-enforced WPs can also emerge in bosonic systems (\eg phonons and photons) due to nonsymmorphic symmetries. We demonstrate that for some nonsymmorphic chiral space groups, several high-symmetry $k$-points can host \emph{only} WPs in the phononic systems, dubbed symmetry-enforced Weyl phonons (SEWPs). The SEWPs, enumerated in Table~\ref{tab:SEWP}, are pinned at the boundary of the three-dimensional (3D) Brillouin zone (BZ) and protected by nonsymmorphic crystal symmetries. By performing first-principles calculations and symmetry analysis, we propose that as an example of SEWPs, the two-fold degeneracies at P are monopole WPs in K$_2$Sn$_2$O$_3$ with space group 199. The two WPs of the same chirality at two nonequivalent P points are related by time-reversal symmetry. In particular, at $\sim 17.5$ THz, a spin-1 Weyl phonon is also found at H, since two Weyl phonons at P carrying a non-zero net Chern number cannot exist alone in the 3D BZ. The significant separation between P and H points makes the surface arcs long and clearly visible. Our findings not only present an effective way to search for WPs in bosonic systems, but also offer some promising candidates for studying monopole Weyl and spin-1 Weyl phonons in realistic materials.
\end{abstract}

\maketitle

\noindent
\textbf{INTRODUCTION}\\
Topological phonons\cite{1,2,3,4,5,6,7,8}, referring to the quantized excited vibrational states of interacting atoms, have been most recently attracted attentions in condensed matter physics because of their unique physical nature\cite{9,10,11,12,13,8}. In similarity to various quasi-particles in electronic systems, topological phonons such as (spin-1/2 or monopole) Weyl, Dirac, spin-1 Weyl and charge-2 Dirac phonons have been predicted/observed in 3D momentum space of solid crystals\cite{15,16,17,18,19,20,21,22,23}, strengthening largely our understanding of elementary particles in the universe. For instance, Zhang \emph{et al.} predicted that both spin-1 Weyl phonons and charge-2 Dirac phonons exist in the CoSi system\cite{23}. 
The coexistence of the above two different classes of topological phononic quasi-particles exhibits exotic topologically nontrivial features, such as noncontractible surface arcs and double-helicoid surface states\cite{24}. Moreover, phonons can be excited to all energy space to generate unusual transport behaviors, since they are not limited by Pauli exclusion principle and Fermi surfaces in materials. 
The phononic systems with these particular properties provide a good platform for studying topological bosonic states in experiments.

\begin{figure}[!th]
\centering
\includegraphics[width=8.5 cm]{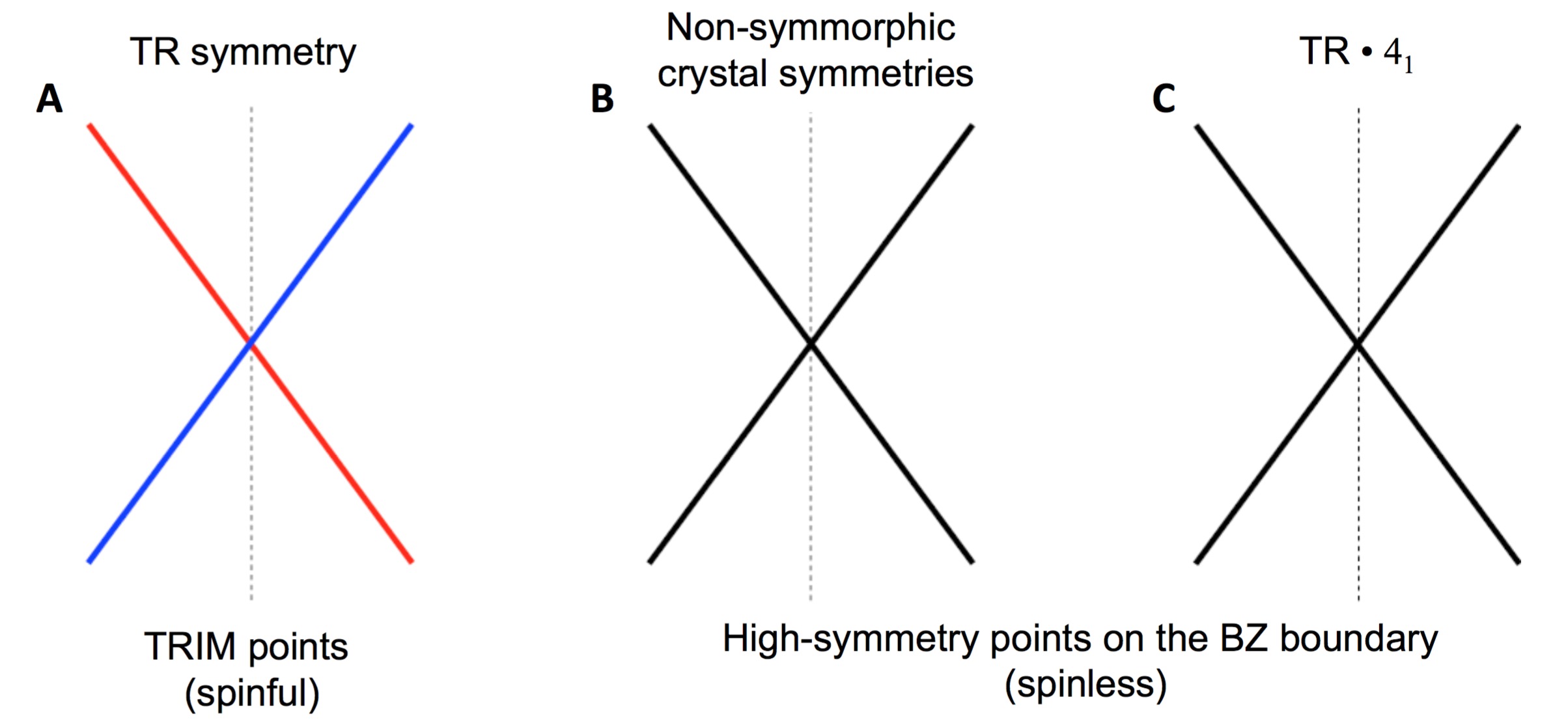}
\caption{\textbf{The schematic symmetry-enforced (spin-1/2 or monopole) Weyl points}. As we know, a WP can be stabilized at a general point in the 3D BZ without any symmetries. However, an additional symmetry can pin the WP at a fixed point. \textbf{A} In a spinful system of a chiral crystal, all the Kramers degenerapcies at the time-reversal invariant momenta (TRIM) are representing WPs. In a spinless system of a chiral crystal, the two-fold degeneracies can also be enforced to form WPs, by some non-symmorphic crystal (unitary) symmetries in \textbf{B} or by the combined (anti-unitary) symmetry of TR symmetry and a four-fold screw symmetry $\bf 4_1$ in \textbf{C}.
}\label{fig:1}
\end{figure}

\begin{table}[!t]
\caption{The list of the SEWPs. The first and the second columns indicate the SG number and the corresponding high-symmetry $k$-point, respectively. The third and fourth columns show the abstract group (AG), which the little group of the $k$-point is isomorphic with, and the corresponding irreps (separated by semicolons). The $k$-point $(uvw)\equiv u\bg_1+v\bg_2+w\bg_3$ and the translation $\{E|abc\}\equiv T(a\bt_1+b\bt_2+c\bt_3)$ are given in units of the primitive reciprocal lattice vectors (\ie $\bg_1,\bg_2,\bg_3$) and primitive lattice vectors (\ie $\bt_1,\bt_2,\bt_3$). See more details for the AGs, their character tables and the definition of these lattice vectors in Ref.~\cite{28}.
}\label{tab:SEWP}
\begin{tabular}{cp{1.1cm}|c|c|c|p{2.2cm}}
\hline
\hline
 space &group & $k$-point & AG & irreps& $\ \ \{A,B\}=0$ \\
\hline
 24 &$I2_12_12_1$ &W $(\frac{1}{4}\frac{1}{4}\frac{1}{4})$ &$G_{16}^7$& R9 & A:$\{C_{2z}|\frac{1}{2}0\frac{1}{2}\}$ B:$\{C_{2y}|\frac{1}{2}\frac{1}{2}0\}$\\
\hline
 80 &$I4_1$ &P $(\frac{1}{4}\frac{1}{4}\frac{1}{4})$ &$G_{2}^1$ & R1R2 & $({\cal T}\cdot{\bf 4_1})^4=-1$ ${\bf 4_1}$:$\{C_{4z}|\frac{3}{4}\frac{1}{4}\frac{1}{2}\}$ \\
\hline
 98 &$I4_122$     &P $(\frac{1}{4}\frac{1}{4}\frac{1}{4})$ &$G_{16}^7$ & R10 & A:$\{C_{2z}|000\}$ B:$\{C_{2y}|0\frac{1}{2}\frac{1}{2}\}$\\
\hline
199 &$I2_13$      &P $(\frac{1}{4}\frac{1}{4}\frac{1}{4})$ &$G_{48}^3$& R7;R8;R9 & A:$\{C_{2x}|\frac{1}{2}\frac{1}{2}0\}$ B:$\{C_{2z}|\frac{3}{2}1\frac{1}{2}\}$\\
\hline
210 &$F4_132$     &W $(\frac{1}{2}\frac{1}{4}\frac{3}{4})$ &$G_{16}^7$& R10 & A:$\{C_{2x}|000\}$ B:$\{C_{2f}|\frac{1}{4}\frac{1}{4}\frac{1}{4}\}$  $C_{2f}$:$xyz \mapsto \bar x\bar z\bar y$\\
\hline
214 &$I4_132$     &P $(\frac{1}{4}\frac{1}{4}\frac{1}{4})$ &$G_{48}^3$& R7;R8;R9 &A:$\{C_{2x}|\frac{1}{2}\frac{1}{2}0\}$ B:$\{C_{2z}|\frac{3}{2}1\frac{1}{2}\}$\\
\hline
\hline
\end{tabular}
\end{table}

According to Nielsen-Ninomiya no-go theorem\cite{25}, WPs always come in pairs with opposite chirality, acting as sources/sinks of Berry curvature in the 3D BZ\cite{26}. The WPs are formed by two bands with linear dispersions in 3D momentum space, which have Chern numbers ($C$) of $\pm 1$ and robust against small perturbations. They are usually not easy to be predicted due to the lack of symmetry protections, and the predictions of WPs usually require comprehensive numerical calculations in the 3D BZ\cite{27}. However, in the spinful electronic systems, Kramers pairs at the time-reversal-invariant momenta are enforced to be WPs by time-reversal (TR) symmetry in chiral crystals (Fig.~\ref{fig:1}A), where there are no improper rotation symmetries, such as inversion or mirror\cite{16,17}. In this work, by checking the symmetries of 230 space groups (SGs), we have uncovered that symmetry-enforced WPs can also emerge in bosonic systems, such as phonons (mainly discussed in the work), photons, and so on, due to the presence of nonsymmorphic symmetries (Figs.~\ref{fig:1}B and C). These WPs are all pinned at the high-symmetry $k$-points on the boundary of the 3D BZ. Unlike the TR-enforced WPs in spinful electronic systems\cite{16}, where the WPs are usually buried in bulk states due to the weak strength of spin-orbit coupling, the nonsymmorphic-crystal-symmetry-enforced Weyl phonons can be well exposed. Consequently, the associated surface arcs are long and robust, which can be easily probed in future experiments. By performing symmetry analysis in 230 SGs in the presence of TR symmetry, we have demonstrated that for some chiral SGs, several high-symmetry \emph{k}-points can host \emph{only} WPs in the phononic systems, dubbed symmetry-enforced Weyl phonons (SEWPs). We enumerate all the SEWPs at the high-symmetry $k$-points of the SGs in Table~\ref{tab:SEWP}. In this Table, all the phonon bands are doubly degenerate at those high-symmetry $k$-points and each two-fold degeneracy represents a WP. Thus, one can easily predict WPs in such systems as long as the materials are of the SGs in Table~\ref{tab:SEWP}. The results significantly lower the difficulty to predict the WPs in bosonic systems.

By employing first-principles calculations, we predicted that as an example of the SEWPs, two-fold degeneracies at the P point are WPs in the crystal of K$_2$Sn$_2$O$_3$ in SG 199. First, there are two non-equivalent P points in the first BZ, which are related by time-reversal symmetry. Therefore, the WPs at two P points host the same chirality. Second, at $\sim$ 17.5 THz, a spin-1 Weyl phonon is also found at H, since two Weyl phonons at P carrying a non-zero net Chern number cannot exist alone in the 3D BZ. Third, the spin-1 Weyl phonon here has been found to locate on the boundary of the BZ, which is robust against the LOTO (longitudinal and transverse optical phonon splitting) modification in the phonon spectrum. Lastly, the symmetry-related WPs host the same chiral charge, giving rise to nontrivial isofrequency surfaces of phonons, associated with \emph{nonzero} Chern numbers.
In addition, the long surface arcs and the double-helicoid states are presented as well.
More examples of SEWPs can be found in the Supplementary Materials (SM). These new findings not only provide an effective way to search for monopole WPs in bosonic systems, but also predict some promising candidates for studying topological quasiparticles in experiments.
~\\£º

\noindent
\textbf{RESULTS AND DISCUSSION}\\
\noindent
\textbf{Searching for SEWPs by symmetry analysis}\\
The guiding principle of our search is to find two-dimensional irreducible representations (irreps) of the (little) group of lattice symmetries at high-symmetry $k$-points in the 3D BZ for each of the 230 SGs in the presence of TR symmetry; the dimension of the irreps corresponds to the number of bands that meet at the high-symmetry $k$-points.
Then, one has to exclude the SGs that contain improper rotational symmetries or two-fold screw symmetries ($\bar C_2$) with $[\bar C_2{\cal T}]^2=-1$, to make sure that there is no double degeneracy along any high-symmetry line/plane crossing these $k$-points.
Since we are interested in bosonic systems, we consider only the single-valued representations; TR symmetry is an antiunitary that squares to 1 (\ie $[{\cal T}]^2=1$).
We find that the two-fold WPs pinned at high-symmetry $k$-points in bosonic systems can have Chern numbers of $\pm 1$ [\ie (monopole) WPs], $\pm 2$ [\ie double WPs] and $\pm 4$ [\ie quadruple WPs], respectively. 
The (monopole) WPs are formed by two bands with linear dispersions\cite{48} and can be stabilized at general points in 3D momentum space, while two bands of the double and quadruple WPs have nonlinear dispersions\cite{48,49,49nat,50,51,52,53}, which are usually pinned at high-symmetry $k$-points by symmetries.
A complete list of the high-symmetry $k$-points, where these three kinds of two-fold WPs can be pinned and protected in bosonic systems, is presented in Table~\ref{tab:SPWP} in Section A of the SM. 
For most of them, besides the two-dimensional irreps, one-dimensional and/or three-dimensional irreps are also allowed.
However, at several specific $k$-points, \emph{only} the two-dimensional irrep of WPs is allowed in the phononic systems, dubbed SEWPs, which we are mainly focused on in this work.

The results of SEWPs are summarized in Table~\ref{tab:SEWP}. These SEWPs are monopole WPs and are located on the boundary of the 3D BZ. All of them are chiral SGs with non-symmorphic symmetries, and all representations are projective; these are in fact necessary ingredients for the (spin-1/2) Weyl excitations in the phononic systems. We find that most of two-fold degeneracies are protected by the anti-commutation relation of two unitary operators (\ie $\{A,B\}=0$) in Table~\ref{tab:SEWP}, except for SG 80. At the P point of SG 80, an antiunitary symmetry of its little group is the combined operator of time reversal ($\cal T$) and four-fold screw symmetry (${\bf 4_1}\equiv\{C_{4z}|\frac{3}{4}\frac{1}{4}\frac{1}{2}\}$)(see the definition in Ref.~\cite{28}. One can check that $({\cal T}\cdot {\bf 4_1})^4=\{E|110\}=e^{2i\pi(\frac{1}{4}+\frac{1}{4})}=-1$, which enforces a Kramers-like degeneracy as discussed in Ref.~\cite{33}.

Then, we take SG 199 as an example to illustrate the anti-commutation relation in the main text (see more derivations for all other SGs in Table~\ref{tab:SEWP} in Section D of the SM). SG 199 hosts only Weyl phonons at the P point (the high-symmetry points are defined in Ref.~\cite{28}), even though it hosts three different irreps of the AG $G_{48}^3$ in Table~\ref{tab:SEWP}. This SG has a body-centered cubic Bravais lattice. The operators $C_{2x}$ and $ C_{2z}$ acting on the primitive lattice vectors ($\bt_1,\bt_2,\bt_3$) are presented\cite{28} as follows:
\begin{equation}
 \begin{split}
& C_{2x} \left(\begin{array}{ccc}
   \bt_1 &\bt_2 &\bt_3  \end{array}\right)=
  \left(\begin{array}{ccc}
   \bt_1 & \bt_2 & \bt_3 \end{array}\right)
  \left[\begin{array}{ccc}
   -1 & 0 & 0 \\
   -1 & 0 & 1 \\
   -1 & 1 & 0 \end{array}\right]; \\
& C_{2z} \left(\begin{array}{ccc}
   \bt_1 &\bt_2 &\bt_3 \end{array}\right)=
  \left(\begin{array}{ccc}
   \bt_1 & \bt_2 & \bt_3 \end{array}\right)
  \left[\begin{array}{ccc}
   0 & 1 & -1 \\
   1 & 0 & -1 \\
   0 & 0 & -1 \end{array}\right].
 \end{split}
\end{equation}
Thus,
\begin{eqnarray}
AB&=& \{C_{2x}|\frac{1}{2}\frac{1}{2}0\}\{C_{2z}|\frac{3}{2}1\frac{1}{2}\}= \{C_{2y}|-1,-\frac{1}{2},-\frac{1}{2}\} \notag\\
BA&=& \{C_{2z}|\frac{3}{2}1\frac{1}{2}\} \{C_{2x}|\frac{1}{2}\frac{1}{2}0\}=\{C_{2y}|2\frac{3}{2}\frac{1}{2}\} \notag \\
&=& \{E|3,2,1\} AB 
\end{eqnarray}
At the P point ($\frac{1}{4},\frac{1}{4},\frac{1}{4}$), the pure translation operator $\{E|3,2,1\}$ is expressed as $e^{2i\pi(3+2+1)/4}=-1$. Therefore, we get $\{A,B\}=0$, which yields all the phonon bands to be at least two-fold degenerate at the P point. Note that no higher $n$-dimensional irreps ($n>2$) are found at P. In addition, we have checked that there is no symmetry-protected degeneracy on the high-symmetry planes/lines crossing the P point.

~\\£º

\noindent
\textbf{Effective $k\cdot p$ models}\\
Let's consider a two-band model at the P point in SG 199 first. We have $A^2=\{E|000\}=1$, $B^2=\{E|220\}=1$ with $E$ the identity operator and $\{A,B\}=0$. With the matrix representations of $A=\sigma_x$ and $B=\sigma_z$, the $k\cdot p$ invariant Hamiltonian is derived as (to the first order),
\begin{equation}
H^{199}_P(\bk)=v_1\sigma_xk_x+v_2\sigma_yk_y+v_3\sigma_zk_z \label{eq:H199}
\end{equation}
with $\sigma_{x,y,z}$ Pauli matrices, $k_{x,y,z}$ momentum offset from P point, and $v_{1,2,3}$ real coefficients. Obviously, it's a Weyl Hamiltonian. Other SGs in Table~\ref{tab:SEWP} with antiunitary commutation relations share the similar results. We notice that the P point is a TR non-invariant point at a corner of the BZ (\ie P $\neq-$P). Those systems host another Weyl phonon of the same chirality at $-$P due to TR symmetry, indicating that there must be some other nontrivial excitation(s) in the systems, since the two Weyl phonons carrying a non-zero net Chern number cannot exist alone in the 3D BZ. Here, we choose K$_2$Sn$_2$O$_3$ in SG 199 as an example for illustration, which exhibits a spin-1 Weyl phonon ($C=2$) at H and two Weyl phonons ($C=-1$) at P between the 39$^{th}$ and 40$^{th}$ bands.

Then we consider the P point in SG 80, where we cannot find two symmetry operators with the anti-commutation relation.
We consider two symmetry operators at P point: $D=\{C_{2z}|100\}={\bf 4}_1^2$ and ${\cal T}\cdot \bf 4_1$, where $\bf 4_1$ is a nonsymmorphic four-fold rotational symmetry, followed by a fractional lattice translation [T($\vec c$/4), where $\vec c$ is a lattice constant in the $z$ direction]. It is worth noting that ${\bf 4_1}$ is not a symmetry operator that keeps P invariant (see more details in the SM). We can express the two operators as $D=-i\sigma_z$ and ${\cal T}\cdot {\bf 4_1}=\frac{1}{\sqrt{2}}(\sigma_x+\sigma_y)\cal K$, which meet the conditions: $({\cal T}\cdot {\bf 4_1})^2=D$ and $({\cal T}\cdot {\bf 4_1})^4=-1$. Thus, the $k\cdot p$ invariant Hamiltonian is derived as (to the first order of $\bk$),
\begin{eqnarray}
H^{80}_P(\bk)&=&(v_1k_x+v_2k_y)\sigma_x+(v_2k_x-v_1k_y)\sigma_y+v_3\sigma_zk_z \notag \\
   &=&v_1[(k_x+\theta k_y)\sigma_x+(\theta k_x-k_y)\sigma_y]+v_3\sigma_zk_z \notag \\
   &=& \sqrt{v_1^2+v_2^2}\left(k_{||}\sigma_x+k_\perp\sigma_y\right)+v_3\sigma_zk_z \label{eq:H80}
\end{eqnarray}
with $\theta=\frac{v_2}{v_1}$, $k_{||}=\frac{k_x+\theta k_y}{\sqrt{1+\theta^2}}$ and $k_\perp=\frac{\theta k_x-k_y}{\sqrt{1+\theta^2}}$.
It's also a Weyl Hamiltonian with isotropy in the $k_x-k_y$ plane.


\begin{figure*}[!tb]
\centering
\includegraphics[width=17.5 cm]{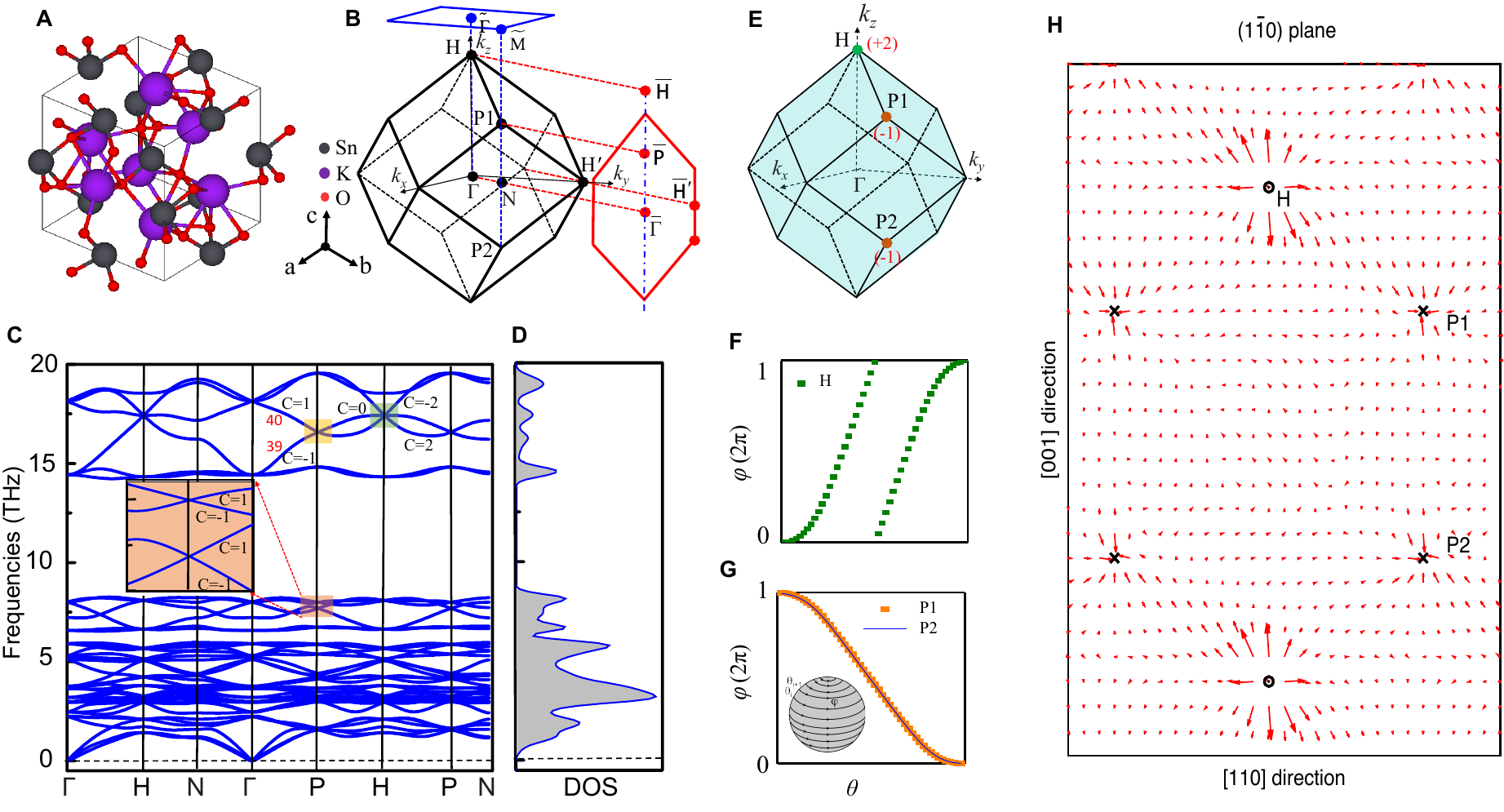}
\caption{\textbf{The crystalline structure of K$_2$Sn$_2$O$_3$ and its phonon dispersions}. \textbf{A} Crystal structure of K$_2$Sn$_2$O$_3$ in a primitive cell, where the purple (black and red) atoms stand for K (Sn and O). \textbf{B} The bulk BZ of the sample and the related (001) and (110) surface BZs. \textbf{C} and \textbf{D} The phonon dispersions of K$_2$Sn$_2$O$_3$ along high-symmetry lines and the density of states (DOS) without the polariton of LOTO splitting. The Chern numbers of some nontrivial phonon branches around P and H points are shown. \textbf{E} The distributions of two monopole WPs (P1 and P2 points) and the spin-1 WP (H point) in the first BZ. \textbf{F} and \textbf{G} The evolutions of Wannier centers of the 39$^{th}$ phonon band for H, P1 and P2. The Wannier centers ($\varphi$) are defined on the series of Wilson loops (parameterized by $\theta$) over a sphere enclosing a WP (inset in \textbf{g}). \textbf{H} The distribution of the Berry curvature in the ($1\bar1 0$) plane. The H point (``o") and the P1/P2 point (``x") are viewed as the source and the sink of Berry curvature, respectively.}\label{fig:2}
\end{figure*}
~\\£º

\noindent
\textbf{SEWPs in realistic materials}\\
Based on our symmetry analysis, the phonon dispersions of any material in the SGs in Table~\ref{tab:SEWP} have to contain WPs at those high-symmetry points. To confirm the theoretical results, we have systematically performed the {\it ab initio} phonon calculations on some materials for each SG in Table~\ref{tab:SEWP}. As an example, we focus on the results and discussions on K$_2$Sn$_2$O$_3$ of SG 199 in the main text, and put the discussions for other SGs in the SM. The crystallographic data of K$_2$Sn$_2$O$_3$ are adopted from Ref.~\cite{34}, and the primitive cell is illustrated in Fig.~\ref{fig:2}A, where the purple (black and red) atoms stand for K (Sn and O) atoms. The material example belongs to the body-centered cubic structure with SG I2$_1$3. Each primitive cell contains 14 atoms with four K, and four Sn and six O atoms. The bulk BZ, (001) surface BZ and (110) surface BZ are shown in Fig.~\ref{fig:2}B.

The calculated phonon dispersions of K$_2$Sn$_2$O$_3$ are shown in Fig.~\ref{fig:2}C. It's clearly seen that there are some band crossings (degeneracies) at high-symmetry $k$-points, especially for the optical dispersions. First, we do find that all the phonon bands are doubly degenerate at P point, resembling monopole WPs. The corresponding results of the Chern number calculations for the bands around P point are shown in Fig.~\ref{fig:2}C and its insets. From the dispersions in Fig.~\ref{fig:2}C, we turn our attention to the WPs formed by the 39$^{th}$ and 40$^{th}$ bands, which have linear dispersions in a wide frequency range. Second, we have also computed the Chern numbers of the two non-equivalent P points (\ie P1 and P2). The chiral charges of WPs at P1 and P2 are computed to be $-1$. Here, the chiral charge of a WP is defined by the Chern number of the lower band, which is computed by the Wilsonloop technique on a sphere enclosing the WP~\cite{46,47}. The results of the sphere for the 39$^{th}$ band around the P point are shown in Fig.~\ref{fig:2}G. It's consistent with TR symmetry in the system, as mentioned before. 
By fitting the two phonon bands in the vicinity of P point, the $v_1,~v_2$ and $v_3$ coefficients in Eq.~(\ref{eq:H199}) are given as 2.19, 2.19 and -2.19 THz$\cdot$\AA.
Third, since two Weyl phonons carrying a non-zero net Chern number cannot exist alone in the 3D BZ, a three-fold spin-1 Weyl phonon is found at H point (highlighted by blue color), formed by the 39$^{th}$, 40$^{th}$ and 41$^{st}$ bands. The Chern numbers of these three bands are computed to be $+2,~0,~-2$, respectively, as shown in Fig.~\ref{fig:2}C. The results of the sphere for the 39$^{th}$ band around the H point are shown in Fig.~\ref{fig:2}F. 
As illustrated Fig.~\ref{fig:2}H, the spin-1 WP with $C=+2$ at H point acts as the ``source" point, whereas two monopole WPs with $C=-1$ at P points can be viewed as the ``sink" points of Berry curvature.

\begin{figure}[!tb]
\centering
\includegraphics[width=8.7 cm]{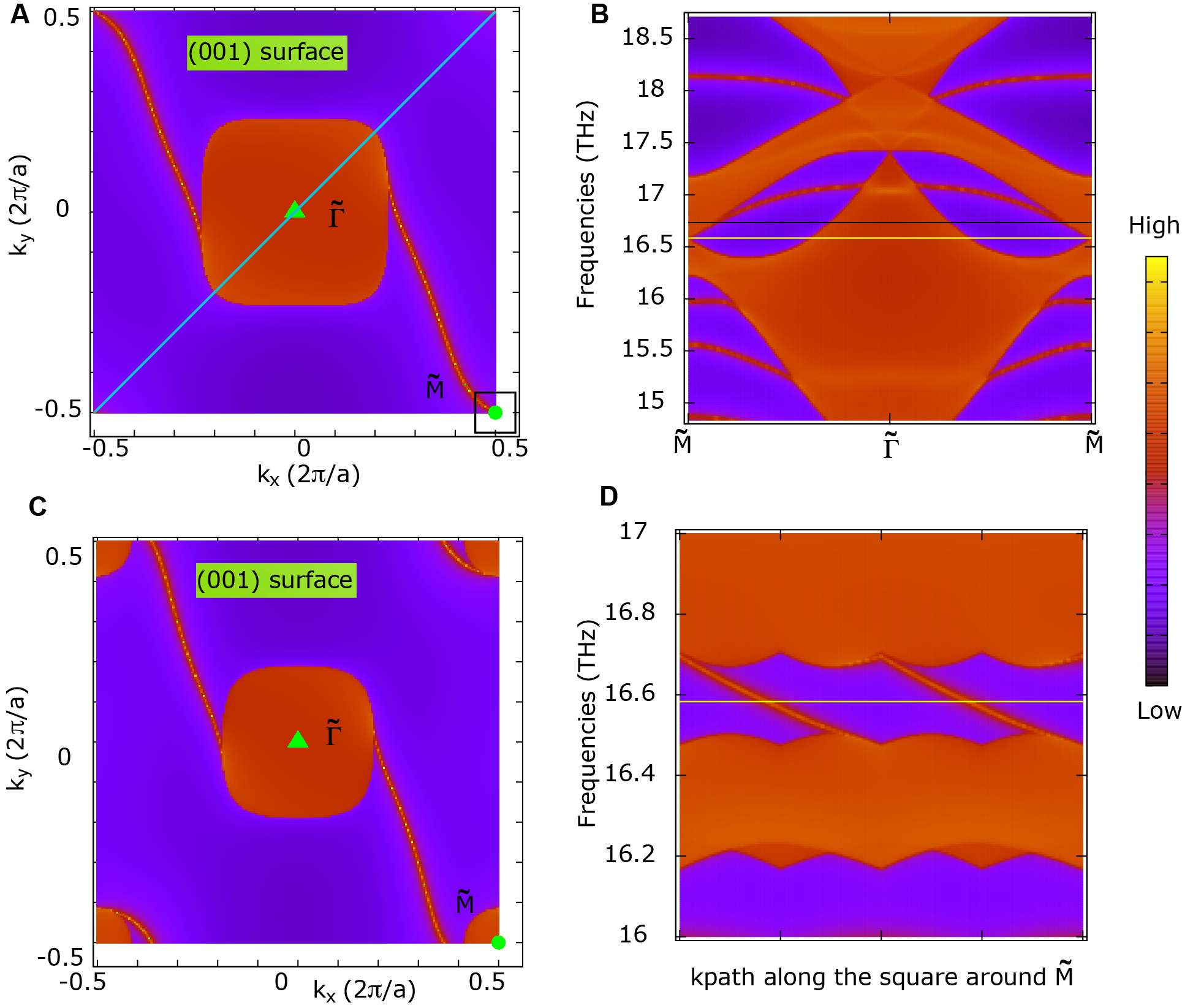}
\caption{\textbf{The (001)-surface phonon dispersions and isofrequency surface contours in K$_2$Sn$_2$O$_3$.} \textbf{A} and \textbf{C} isofrequency surface contours at two frequencies, 16.58 and 16.73 THz. \textbf{B} The surface phonon dispersions along $\widetilde{\textrm{M}}$-$\widetilde{\Gamma}$-$\widetilde{\textrm{M}}$. The two frequencies are indicated by two horizontal lines in \textbf{B}. \textbf{D} The surface phonon dispersions along the square path around the $\widetilde{\textrm{M}}$ point in \textbf{A}, which demonstrate the double-helical surface states.} \label{fig:3}
\end{figure}

Consequently, this system hosts several unconventional properties: (i) The symmetry-related WPs are of the same chirality, \ie the monopole WPs at the P points, which can generate the nontrivial isofrequency surfaces of photon with \emph{nonzero} Chern numbers. (ii) The nontrivial excitations (quasiparticles) are pinned at the high-symmetry $k$-points, such as the monopole Weyl phonons at P and the three-fold spin-1 Weyl phonon at H, which give rise to large surface arcs due to the large separation of the sources and sinks of the Berry curvature. (iii) The spin-1 Weyl phonon is proposed on the boundary of the BZ of a realistic material and is robust against the LOTO modification (see Fig. S1 in the SM), supporting that it could be observed in future experiments. Such a spin-1 Weyl phonon at H is also found in K$_8$ carbon of SG 214 in Section D of the SM.
~\\£º

\noindent
\textbf{Exotic surface states}\\
Then, we turn to examine the isofrequency surface contours and the surface arcs to explore the exotic physical behaviors of the SEWPs. As the two P points are projected to the BZ corner ($\widetilde{\textrm{M}}$) and the H point to the BZ center ($\widetilde{\Gamma}$) in Fig.~\ref{fig:3}A, we plot the (001)-surface phonon dispersions along the green line of Fig.~\ref{fig:3}A (\ie $\widetilde{\textrm{M}}$-$\widetilde{\Gamma}$-$\widetilde{\textrm{M}}$) in Fig.~\ref{fig:3}B. At two chosen frequencies of 16.58 (\ie the frequency of the WPs at P) and 16.73 THz, the arc-like surface states are illustrated in Figs.~\ref{fig:3}A and~\ref{fig:3}C, respectively. 
The surface arcs connect the two monopole WPs at $\widetilde{\textrm{M}}$ and one spin-1 Weyl phonon at $\widetilde{\Gamma}$.
As expected, the isofrequency surface contours display double-helicoid states\cite{36}, which are spiral clockwise with increasing frequency. 
The surface phonon dispersions along the square-shaped path around $\widetilde{\textrm{M}}$ show double-helicoid surface states (see Fig.~\ref{fig:3}D), which verifies further the topological nontrivial feature of the two monopole WPs at P points.

\begin{figure}[!tb]
\centering
\includegraphics[width=8.6 cm]{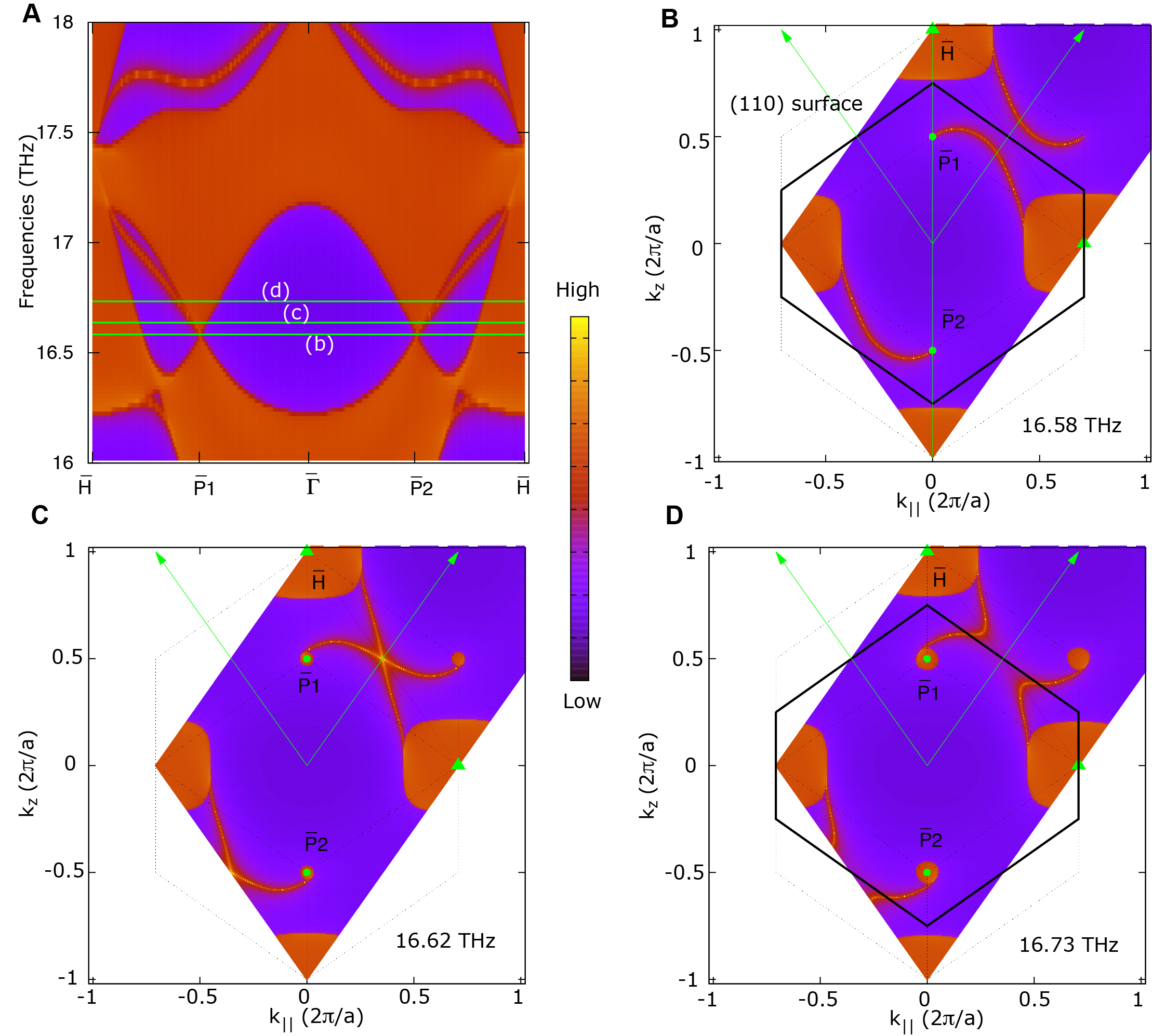}
\caption{\textbf{The (110)-surface phonon dispersions and isofrequency surface contours in K$_2$Sn$_2$O$_3$.} 
\textbf{A} The surface phonon dispersions along the vertical line (\ie $\overline{\textrm{H}}$-$\overline{P}1$-$\overline{\Gamma}$-$\overline{P}2$-$\overline{\textrm{H}}$) on the (110)-surface BZ of Fig.~\ref{fig:1}B.  \textbf{B}-\textbf{D} The isofrequency surface contours at three frequences, 16.58, 16.62 and 16.73 THz as indicated by the three horizontal lines in \textbf{A}. One can find that Lifshitz transition occurs in the surface arc states in the frequency range.
} \label{fig:4}
\end{figure}

Next, we plot the (110)-surface phonon dispersions in Fig.~\ref{fig:4}A and the isofrequency surface contours in Figs.~\ref{fig:4}B-\ref{fig:4}D.  On the (110)-surface BZ (Fig.~\ref{fig:1}B), the H, P1 and P2 points are projected to $\overline{\textrm{H}}$, $\overline{\textrm{P}}$1 and $\overline{\textrm{P}}$2, respectively. The evolution of the surface arcs is illustrated by isofrequency surface contours for three frequencies, in which the surface arcs are clearly visualized. In this frequency range, a Lifshitz transition occurs clearly in the arc-like surface states. Moreover, the surface arcs at higher frequencies are presented in Fig. S2 of the SM.
~\\£º

\noindent
\textbf{Discussion}\\
By performing symmetry analysis in 230 SGs in the presence of TR symmetry, we demonstrate that there are also symmetry-enforced WPs in the bosonic systems, \eg phonons. We have given a complete list of high-symmetry $k$-points, where two-fold Weyl nodes (\eg $C=\pm 1,~\pm 2$, and $\pm 4$) are protected. This list can guide future experiments in the study of various WPs. Among them, several $k$-points can support \emph{only} two-fold Weyl phonons, dubbed SEWPs. Such SEWPs have Chern numbers of $\pm 1$ and appear in the nonsymmorphic chiral crystals due to the lack of improper rotation symmetries and the presence of nonsymmorphic symmetries. We summarize the high-symmetry $k$-points of SEWPs in Table~\ref{tab:SEWP}. The SEWPs are pinned at the boundary of the 3D BZ by nonsymmorphic symmetries, or the combined symmetry of TR and ${\bf 4_1}$. To confirm our findings, we have systematically investigated the phonon spectra of some realistic materials of the SGs in Table~\ref{tab:SEWP} by using first-principles calculations. Taking K$_2$Sn$_2$O$_3$ of SG 199 as an example, two Weyl phonons at P and a spin-1 Weyl phonon at H appear together in between the 39$^{th}$ and 40$^{th}$ bands. The corresponding surface phonon dispersions display multiple double-helicoid surface states. In addition, the significant separation between the points P and H forms very long and visible surface arcs. Our findings not only present an effective way to search for WPs in bosonic systems, but also offer some promising candidates for studying monopole Weyl and spin-1 Weyl phonons.

~\\£º
\noindent
\textbf{METHODS}\\
\noindent
We carried out the density functional theory (DFT) calculations using the Vienna \emph{ab initio} Simulation Package (VASP)\cite{37,38,39} with the generalized gradient approximation (GGA) in the form of Pardew-Burke-Ernzerhof (PBE) function for the exchange-correlation potential\cite{40,41,42}. An accurate optimization of structural parameters is employed by minimizing the interionic forces below $10^{-6}$$\textrm{eV}/\textrm{{\AA}}$ and an energy cut off at 520 eV. The BZ is gridded with 3$\times$3$\times$3 \emph{k} points. Then the phonon dispersions are gained using the density functional perturbation theory (DFPT), implemented in the Phonopy Package\cite{43}. The force constants are calculated using a $2\times2\times2$ supercell. To reveal the phonon topological nature, we constructed the phononic Hamiltonian of the tight-binding (TB) model and obtained the surface local density of states (LDOS) with the open-source software Wanniertools\cite{44} code and surface Green's functions\cite{45}. The irreps of the phonon bands can be computed by the program --- $ir2tb$ --- on the phononic Hamiltonin of the TB model\cite{gao2020irvsp}. Wilson loop method\cite{46,47} is used to find the Chern numbers or topological charge of monopole WPs and spin-1 WPs.
~\\£º

~\\£º
\noindent
\textbf{DATA AVAILABILITY}\\
Data are available from the authors upon reasonable request.

~\\£º
\noindent
\textbf{ACKNOWLEDGEMENTS}\\
This work is supported by the National Natural Science Foundation of China (Grants No. 11774104, 11504117 and 11274128) and 
the Strategic Priority Research Program of Chinese Academy of Sciences (Grant No. XDB33000000).

~\\£º
\noindent
\textbf{AUTHOR CONTRIBUTIONS}\\
Z.W. and H.H.F. conceived and supervised the project. Q.B.L. and H.H. F. did the phonon calculations. Y.T.Q. and Z.W. did the symmetry analysis. All authors contributed to analysing the results and writing the manuscript.




\bibliography{tps}

\begin{thebibliography}{53}
\expandafter\ifx\csname natexlab\endcsname\relax\def\natexlab#1{#1}\fi
\expandafter\ifx\csname bibnamefont\endcsname\relax
  \def\bibnamefont#1{#1}\fi
\expandafter\ifx\csname bibfnamefont\endcsname\relax
  \def\bibfnamefont#1{#1}\fi
\expandafter\ifx\csname citenamefont\endcsname\relax
  \def\citenamefont#1{#1}\fi
\expandafter\ifx\csname url\endcsname\relax
  \def\url#1{\texttt{#1}}\fi
\expandafter\ifx\csname urlprefix\endcsname\relax\def\urlprefix{URL }\fi
\providecommand{\bibinfo}[2]{#2}
\providecommand{\eprint}[2][]{\url{#2}}

\bibitem[{\citenamefont{Kane and Lubensky}(2014)}]{1}
\bibinfo{author}{\bibfnamefont{C.}~\bibnamefont{Kane}} \bibnamefont{and}
  \bibinfo{author}{\bibfnamefont{T.}~\bibnamefont{Lubensky}},
  \bibinfo{journal}{Nature Physics} \textbf{\bibinfo{volume}{10}},
  \bibinfo{pages}{39} (\bibinfo{year}{2014}).

\bibitem[{\citenamefont{Prodan and Prodan}(2009)}]{2}
\bibinfo{author}{\bibfnamefont{E.}~\bibnamefont{Prodan}} \bibnamefont{and}
  \bibinfo{author}{\bibfnamefont{C.}~\bibnamefont{Prodan}},
  \bibinfo{journal}{Physical review letters} \textbf{\bibinfo{volume}{103}},
  \bibinfo{pages}{248101} (\bibinfo{year}{2009}).

\bibitem[{\citenamefont{Zhang et~al.}(2010)\citenamefont{Zhang, Ren, Wang, and
  Li}}]{3}
\bibinfo{author}{\bibfnamefont{L.}~\bibnamefont{Zhang}},
  \bibinfo{author}{\bibfnamefont{J.}~\bibnamefont{Ren}},
  \bibinfo{author}{\bibfnamefont{J.-S.} \bibnamefont{Wang}}, \bibnamefont{and}
  \bibinfo{author}{\bibfnamefont{B.}~\bibnamefont{Li}},
  \bibinfo{journal}{Physical review letters} \textbf{\bibinfo{volume}{105}},
  \bibinfo{pages}{225901} (\bibinfo{year}{2010}).

\bibitem[{\citenamefont{Liu et~al.}(2017)\citenamefont{Liu, Xu, Zhang, and
  Duan}}]{4}
\bibinfo{author}{\bibfnamefont{Y.}~\bibnamefont{Liu}},
  \bibinfo{author}{\bibfnamefont{Y.}~\bibnamefont{Xu}},
  \bibinfo{author}{\bibfnamefont{S.-C.} \bibnamefont{Zhang}}, \bibnamefont{and}
  \bibinfo{author}{\bibfnamefont{W.}~\bibnamefont{Duan}},
  \bibinfo{journal}{Physical Review B} \textbf{\bibinfo{volume}{96}},
  \bibinfo{pages}{064106} (\bibinfo{year}{2017}).

\bibitem[{\citenamefont{Chen et~al.}(2014)\citenamefont{Chen, Upadhyaya, and
  Vitelli}}]{5}
\bibinfo{author}{\bibfnamefont{B.~G.-g.} \bibnamefont{Chen}},
  \bibinfo{author}{\bibfnamefont{N.}~\bibnamefont{Upadhyaya}},
  \bibnamefont{and} \bibinfo{author}{\bibfnamefont{V.}~\bibnamefont{Vitelli}},
  \bibinfo{journal}{Proceedings of the National Academy of Sciences}
  \textbf{\bibinfo{volume}{111}}, \bibinfo{pages}{13004}
  (\bibinfo{year}{2014}).

\bibitem[{\citenamefont{Wang et~al.}(2015)\citenamefont{Wang, Lu, and
  Bertoldi}}]{6}
\bibinfo{author}{\bibfnamefont{P.}~\bibnamefont{Wang}},
  \bibinfo{author}{\bibfnamefont{L.}~\bibnamefont{Lu}}, \bibnamefont{and}
  \bibinfo{author}{\bibfnamefont{K.}~\bibnamefont{Bertoldi}},
  \bibinfo{journal}{Physical review letters} \textbf{\bibinfo{volume}{115}},
  \bibinfo{pages}{104302} (\bibinfo{year}{2015}).

\bibitem[{\citenamefont{S{\"u}sstrunk and Huber}(2015)}]{7}
\bibinfo{author}{\bibfnamefont{R.}~\bibnamefont{S{\"u}sstrunk}}
  \bibnamefont{and} \bibinfo{author}{\bibfnamefont{S.~D.} \bibnamefont{Huber}},
  \bibinfo{journal}{Science} \textbf{\bibinfo{volume}{349}},
  \bibinfo{pages}{47} (\bibinfo{year}{2015}).

\bibitem[{\citenamefont{Mousavi et~al.}(2015)\citenamefont{Mousavi, Khanikaev,
  and Wang}}]{8}
\bibinfo{author}{\bibfnamefont{S.~H.} \bibnamefont{Mousavi}},
  \bibinfo{author}{\bibfnamefont{A.~B.} \bibnamefont{Khanikaev}},
  \bibnamefont{and} \bibinfo{author}{\bibfnamefont{Z.}~\bibnamefont{Wang}},
  \bibinfo{journal}{Nat. Commun.} \textbf{\bibinfo{volume}{6}},
  \bibinfo{pages}{8682} (\bibinfo{year}{2015}).

\bibitem[{\citenamefont{He et~al.}(2016)\citenamefont{He, Ni, Ge, Sun, Chen,
  Lu, Liu, and Chen}}]{9}
\bibinfo{author}{\bibfnamefont{C.}~\bibnamefont{He}},
  \bibinfo{author}{\bibfnamefont{X.}~\bibnamefont{Ni}},
  \bibinfo{author}{\bibfnamefont{H.}~\bibnamefont{Ge}},
  \bibinfo{author}{\bibfnamefont{X.-C.} \bibnamefont{Sun}},
  \bibinfo{author}{\bibfnamefont{Y.-B.} \bibnamefont{Chen}},
  \bibinfo{author}{\bibfnamefont{M.-H.} \bibnamefont{Lu}},
  \bibinfo{author}{\bibfnamefont{X.-P.} \bibnamefont{Liu}}, \bibnamefont{and}
  \bibinfo{author}{\bibfnamefont{Y.-F.} \bibnamefont{Chen}},
  \bibinfo{journal}{Nature physics} \textbf{\bibinfo{volume}{12}},
  \bibinfo{pages}{1124} (\bibinfo{year}{2016}).

\bibitem[{\citenamefont{Lu et~al.}(2013)\citenamefont{Lu, Fu, Joannopoulos, and
  Solja{\v{c}}i{\'c}}}]{10}
\bibinfo{author}{\bibfnamefont{L.}~\bibnamefont{Lu}},
  \bibinfo{author}{\bibfnamefont{L.}~\bibnamefont{Fu}},
  \bibinfo{author}{\bibfnamefont{J.~D.} \bibnamefont{Joannopoulos}},
  \bibnamefont{and}
  \bibinfo{author}{\bibfnamefont{M.}~\bibnamefont{Solja{\v{c}}i{\'c}}},
  \bibinfo{journal}{Nature photonics} \textbf{\bibinfo{volume}{7}},
  \bibinfo{pages}{294} (\bibinfo{year}{2013}).

\bibitem[{\citenamefont{Lu et~al.}(2015)\citenamefont{Lu, Wang, Ye, Ran, Fu,
  Joannopoulos, and Solja{\v{c}}i{\'c}}}]{11}
\bibinfo{author}{\bibfnamefont{L.}~\bibnamefont{Lu}},
  \bibinfo{author}{\bibfnamefont{Z.}~\bibnamefont{Wang}},
  \bibinfo{author}{\bibfnamefont{D.}~\bibnamefont{Ye}},
  \bibinfo{author}{\bibfnamefont{L.}~\bibnamefont{Ran}},
  \bibinfo{author}{\bibfnamefont{L.}~\bibnamefont{Fu}},
  \bibinfo{author}{\bibfnamefont{J.~D.} \bibnamefont{Joannopoulos}},
  \bibnamefont{and}
  \bibinfo{author}{\bibfnamefont{M.}~\bibnamefont{Solja{\v{c}}i{\'c}}},
  \bibinfo{journal}{Science} \textbf{\bibinfo{volume}{349}},
  \bibinfo{pages}{622} (\bibinfo{year}{2015}).

\bibitem[{\citenamefont{Zhang and Niu}(2015)}]{12}
\bibinfo{author}{\bibfnamefont{L.}~\bibnamefont{Zhang}} \bibnamefont{and}
  \bibinfo{author}{\bibfnamefont{Q.}~\bibnamefont{Niu}},
  \bibinfo{journal}{Physical review letters} \textbf{\bibinfo{volume}{115}},
  \bibinfo{pages}{115502} (\bibinfo{year}{2015}).

\bibitem[{\citenamefont{Gao et~al.}(2018)\citenamefont{Gao, Zhang, and
  Zhang}}]{13}
\bibinfo{author}{\bibfnamefont{M.}~\bibnamefont{Gao}},
  \bibinfo{author}{\bibfnamefont{W.}~\bibnamefont{Zhang}}, \bibnamefont{and}
  \bibinfo{author}{\bibfnamefont{L.}~\bibnamefont{Zhang}},
  \bibinfo{journal}{Nano letters} \textbf{\bibinfo{volume}{18}},
  \bibinfo{pages}{4424} (\bibinfo{year}{2018}).

\bibitem[{\citenamefont{Stenull et~al.}(2016)\citenamefont{Stenull, Kane, and
  Lubensky}}]{15}
\bibinfo{author}{\bibfnamefont{O.}~\bibnamefont{Stenull}},
  \bibinfo{author}{\bibfnamefont{C.}~\bibnamefont{Kane}}, \bibnamefont{and}
  \bibinfo{author}{\bibfnamefont{T.}~\bibnamefont{Lubensky}},
  \bibinfo{journal}{Physical review letters} \textbf{\bibinfo{volume}{117}},
  \bibinfo{pages}{068001} (\bibinfo{year}{2016}).

\bibitem[{\citenamefont{Chang et~al.}(2018)\citenamefont{Chang, Wieder,
  Schindler, Sanchez, Belopolski, Huang, Singh, Wu, Chang, Neupert
  et~al.}}]{16}
\bibinfo{author}{\bibfnamefont{G.}~\bibnamefont{Chang}},
  \bibinfo{author}{\bibfnamefont{B.~J.} \bibnamefont{Wieder}},
  \bibinfo{author}{\bibfnamefont{F.}~\bibnamefont{Schindler}},
  \bibinfo{author}{\bibfnamefont{D.~S.} \bibnamefont{Sanchez}},
  \bibinfo{author}{\bibfnamefont{I.}~\bibnamefont{Belopolski}},
  \bibinfo{author}{\bibfnamefont{S.-M.} \bibnamefont{Huang}},
  \bibinfo{author}{\bibfnamefont{B.}~\bibnamefont{Singh}},
  \bibinfo{author}{\bibfnamefont{D.}~\bibnamefont{Wu}},
  \bibinfo{author}{\bibfnamefont{T.-R.} \bibnamefont{Chang}},
  \bibinfo{author}{\bibfnamefont{T.}~\bibnamefont{Neupert}},
  \bibnamefont{et~al.}, \bibinfo{journal}{Nature materials}
  \textbf{\bibinfo{volume}{17}}, \bibinfo{pages}{978} (\bibinfo{year}{2018}).

\bibitem[{\citenamefont{Chang et~al.}(2017)\citenamefont{Chang, Xu, Wieder,
  Sanchez, Huang, Belopolski, Chang, Zhang, Bansil, Lin et~al.}}]{17}
\bibinfo{author}{\bibfnamefont{G.}~\bibnamefont{Chang}},
  \bibinfo{author}{\bibfnamefont{S.-Y.} \bibnamefont{Xu}},
  \bibinfo{author}{\bibfnamefont{B.~J.} \bibnamefont{Wieder}},
  \bibinfo{author}{\bibfnamefont{D.~S.} \bibnamefont{Sanchez}},
  \bibinfo{author}{\bibfnamefont{S.-M.} \bibnamefont{Huang}},
  \bibinfo{author}{\bibfnamefont{I.}~\bibnamefont{Belopolski}},
  \bibinfo{author}{\bibfnamefont{T.-R.} \bibnamefont{Chang}},
  \bibinfo{author}{\bibfnamefont{S.}~\bibnamefont{Zhang}},
  \bibinfo{author}{\bibfnamefont{A.}~\bibnamefont{Bansil}},
  \bibinfo{author}{\bibfnamefont{H.}~\bibnamefont{Lin}}, \bibnamefont{et~al.},
  \bibinfo{journal}{Physical review letters} \textbf{\bibinfo{volume}{119}},
  \bibinfo{pages}{206401} (\bibinfo{year}{2017}).

\bibitem[{\citenamefont{Liu et~al.}(2019)\citenamefont{Liu, Fu, Xu, Yu, and
  Wu}}]{18}
\bibinfo{author}{\bibfnamefont{Q.-B.} \bibnamefont{Liu}},
  \bibinfo{author}{\bibfnamefont{H.-H.} \bibnamefont{Fu}},
  \bibinfo{author}{\bibfnamefont{G.}~\bibnamefont{Xu}},
  \bibinfo{author}{\bibfnamefont{R.}~\bibnamefont{Yu}}, \bibnamefont{and}
  \bibinfo{author}{\bibfnamefont{R.}~\bibnamefont{Wu}}, \bibinfo{journal}{The
  journal of physical chemistry letters} \textbf{\bibinfo{volume}{10}},
  \bibinfo{pages}{4045} (\bibinfo{year}{2019}).

\bibitem[{\citenamefont{Xie et~al.}(2019{\natexlab{a}})\citenamefont{Xie, Liu,
  Cheng, Liu, Chen, and Tian}}]{19}
\bibinfo{author}{\bibfnamefont{B.}~\bibnamefont{Xie}},
  \bibinfo{author}{\bibfnamefont{H.}~\bibnamefont{Liu}},
  \bibinfo{author}{\bibfnamefont{H.}~\bibnamefont{Cheng}},
  \bibinfo{author}{\bibfnamefont{Z.}~\bibnamefont{Liu}},
  \bibinfo{author}{\bibfnamefont{S.}~\bibnamefont{Chen}}, \bibnamefont{and}
  \bibinfo{author}{\bibfnamefont{J.}~\bibnamefont{Tian}},
  \bibinfo{journal}{Physical review letters} \textbf{\bibinfo{volume}{122}},
  \bibinfo{pages}{104302} (\bibinfo{year}{2019}{\natexlab{a}}).

\bibitem[{\citenamefont{Yang et~al.}(2019)\citenamefont{Yang, Sun, Xia, Xue,
  Gao, Ge, Jia, Yuan, Chong, and Zhang}}]{20}
\bibinfo{author}{\bibfnamefont{Y.}~\bibnamefont{Yang}},
  \bibinfo{author}{\bibfnamefont{H.-x.} \bibnamefont{Sun}},
  \bibinfo{author}{\bibfnamefont{J.-p.} \bibnamefont{Xia}},
  \bibinfo{author}{\bibfnamefont{H.}~\bibnamefont{Xue}},
  \bibinfo{author}{\bibfnamefont{Z.}~\bibnamefont{Gao}},
  \bibinfo{author}{\bibfnamefont{Y.}~\bibnamefont{Ge}},
  \bibinfo{author}{\bibfnamefont{D.}~\bibnamefont{Jia}},
  \bibinfo{author}{\bibfnamefont{S.-q.} \bibnamefont{Yuan}},
  \bibinfo{author}{\bibfnamefont{Y.}~\bibnamefont{Chong}}, \bibnamefont{and}
  \bibinfo{author}{\bibfnamefont{B.}~\bibnamefont{Zhang}},
  \bibinfo{journal}{Nature Physics} \textbf{\bibinfo{volume}{15}},
  \bibinfo{pages}{645} (\bibinfo{year}{2019}).

\bibitem[{\citenamefont{Xia et~al.}(2019)\citenamefont{Xia, Wang, Chen, Zhao,
  and Xu}}]{21}
\bibinfo{author}{\bibfnamefont{B.}~\bibnamefont{Xia}},
  \bibinfo{author}{\bibfnamefont{R.}~\bibnamefont{Wang}},
  \bibinfo{author}{\bibfnamefont{Z.}~\bibnamefont{Chen}},
  \bibinfo{author}{\bibfnamefont{Y.}~\bibnamefont{Zhao}}, \bibnamefont{and}
  \bibinfo{author}{\bibfnamefont{H.}~\bibnamefont{Xu}},
  \bibinfo{journal}{Physical review letters} \textbf{\bibinfo{volume}{123}},
  \bibinfo{pages}{065501} (\bibinfo{year}{2019}).

\bibitem[{\citenamefont{Li et~al.}(2018)\citenamefont{Li, Huang, Lu, Ma, and
  Liu}}]{22}
\bibinfo{author}{\bibfnamefont{F.}~\bibnamefont{Li}},
  \bibinfo{author}{\bibfnamefont{X.}~\bibnamefont{Huang}},
  \bibinfo{author}{\bibfnamefont{J.}~\bibnamefont{Lu}},
  \bibinfo{author}{\bibfnamefont{J.}~\bibnamefont{Ma}}, \bibnamefont{and}
  \bibinfo{author}{\bibfnamefont{Z.}~\bibnamefont{Liu}},
  \bibinfo{journal}{Nature Physics} \textbf{\bibinfo{volume}{14}},
  \bibinfo{pages}{30} (\bibinfo{year}{2018}).

\bibitem[{\citenamefont{Zhang et~al.}(2018)\citenamefont{Zhang, Song,
  Alexandradinata, Weng, Fang, Lu, and Fang}}]{23}
\bibinfo{author}{\bibfnamefont{T.}~\bibnamefont{Zhang}},
  \bibinfo{author}{\bibfnamefont{Z.}~\bibnamefont{Song}},
  \bibinfo{author}{\bibfnamefont{A.}~\bibnamefont{Alexandradinata}},
  \bibinfo{author}{\bibfnamefont{H.}~\bibnamefont{Weng}},
  \bibinfo{author}{\bibfnamefont{C.}~\bibnamefont{Fang}},
  \bibinfo{author}{\bibfnamefont{L.}~\bibnamefont{Lu}}, \bibnamefont{and}
  \bibinfo{author}{\bibfnamefont{Z.}~\bibnamefont{Fang}},
  \bibinfo{journal}{Physical review letters} \textbf{\bibinfo{volume}{120}},
  \bibinfo{pages}{016401} (\bibinfo{year}{2018}).

\bibitem[{\citenamefont{Miao et~al.}(2018)\citenamefont{Miao, Zhang, Wang,
  Meyers, Said, Wang, Shi, Weng, Fang, and Dean}}]{24}
\bibinfo{author}{\bibfnamefont{H.}~\bibnamefont{Miao}},
  \bibinfo{author}{\bibfnamefont{T.}~\bibnamefont{Zhang}},
  \bibinfo{author}{\bibfnamefont{L.}~\bibnamefont{Wang}},
  \bibinfo{author}{\bibfnamefont{D.}~\bibnamefont{Meyers}},
  \bibinfo{author}{\bibfnamefont{A.}~\bibnamefont{Said}},
  \bibinfo{author}{\bibfnamefont{Y.}~\bibnamefont{Wang}},
  \bibinfo{author}{\bibfnamefont{Y.}~\bibnamefont{Shi}},
  \bibinfo{author}{\bibfnamefont{H.}~\bibnamefont{Weng}},
  \bibinfo{author}{\bibfnamefont{Z.}~\bibnamefont{Fang}}, \bibnamefont{and}
  \bibinfo{author}{\bibfnamefont{M.}~\bibnamefont{Dean}},
  \bibinfo{journal}{Physical Review Letters} \textbf{\bibinfo{volume}{121}},
  \bibinfo{pages}{035302} (\bibinfo{year}{2018}).

\bibitem[{\citenamefont{Bradley and Cracknell}(2009)}]{28}
\bibinfo{author}{\bibfnamefont{C.}~\bibnamefont{Bradley}} \bibnamefont{and}
  \bibinfo{author}{\bibfnamefont{A.}~\bibnamefont{Cracknell}},
  \emph{\bibinfo{title}{The mathematical theory of symmetry in solids:
  representation theory for point groups and space groups}}
  (\bibinfo{publisher}{Oxford University Press}, \bibinfo{year}{2009}).

\bibitem[{\citenamefont{Nielsen and Ninomiya}(1981)}]{25}
\bibinfo{author}{\bibfnamefont{H.~B.} \bibnamefont{Nielsen}} \bibnamefont{and}
  \bibinfo{author}{\bibfnamefont{M.}~\bibnamefont{Ninomiya}},
  \bibinfo{journal}{Nuclear Physics B} \textbf{\bibinfo{volume}{185}},
  \bibinfo{pages}{20} (\bibinfo{year}{1981}).

\bibitem[{\citenamefont{Xie et~al.}(2019{\natexlab{b}})\citenamefont{Xie, Li,
  Ullah, Li, Wang, Li, Li, Yunoki, and Chen}}]{26}
\bibinfo{author}{\bibfnamefont{Q.}~\bibnamefont{Xie}},
  \bibinfo{author}{\bibfnamefont{J.}~\bibnamefont{Li}},
  \bibinfo{author}{\bibfnamefont{S.}~\bibnamefont{Ullah}},
  \bibinfo{author}{\bibfnamefont{R.}~\bibnamefont{Li}},
  \bibinfo{author}{\bibfnamefont{L.}~\bibnamefont{Wang}},
  \bibinfo{author}{\bibfnamefont{D.}~\bibnamefont{Li}},
  \bibinfo{author}{\bibfnamefont{Y.}~\bibnamefont{Li}},
  \bibinfo{author}{\bibfnamefont{S.}~\bibnamefont{Yunoki}}, \bibnamefont{and}
  \bibinfo{author}{\bibfnamefont{X.-Q.} \bibnamefont{Chen}},
  \bibinfo{journal}{Physical Review B} \textbf{\bibinfo{volume}{99}},
  \bibinfo{pages}{174306} (\bibinfo{year}{2019}{\natexlab{b}}).

\bibitem[{\citenamefont{Armitage et~al.}(2018)\citenamefont{Armitage, Mele, and
  Vishwanath}}]{27}
\bibinfo{author}{\bibfnamefont{N.}~\bibnamefont{Armitage}},
  \bibinfo{author}{\bibfnamefont{E.}~\bibnamefont{Mele}}, \bibnamefont{and}
  \bibinfo{author}{\bibfnamefont{A.}~\bibnamefont{Vishwanath}},
  \bibinfo{journal}{Reviews of Modern Physics} \textbf{\bibinfo{volume}{90}},
  \bibinfo{pages}{015001} (\bibinfo{year}{2018}).

\bibitem[{\citenamefont{Wang et~al.}(2020)\citenamefont{Wang, Xia, Chen, Zheng,
  Zhao, and Xu}}]{48}
\bibinfo{author}{\bibfnamefont{R.}~\bibnamefont{Wang}},
  \bibinfo{author}{\bibfnamefont{B.}~\bibnamefont{Xia}},
  \bibinfo{author}{\bibfnamefont{Z.}~\bibnamefont{Chen}},
  \bibinfo{author}{\bibfnamefont{B.}~\bibnamefont{Zheng}},
  \bibinfo{author}{\bibfnamefont{Y.}~\bibnamefont{Zhao}}, \bibnamefont{and}
  \bibinfo{author}{\bibfnamefont{H.}~\bibnamefont{Xu}},
  \bibinfo{journal}{Physical Review Letters} \textbf{\bibinfo{volume}{124}},
  \bibinfo{pages}{105303} (\bibinfo{year}{2020}).

\bibitem[{\citenamefont{He et~al.}(2020)\citenamefont{He, Qiu, Cai, Xiao, Ke,
  Zhang, and Liu}}]{49}
\bibinfo{author}{\bibfnamefont{H.}~\bibnamefont{He}},
  \bibinfo{author}{\bibfnamefont{C.}~\bibnamefont{Qiu}},
  \bibinfo{author}{\bibfnamefont{X.}~\bibnamefont{Cai}},
  \bibinfo{author}{\bibfnamefont{M.}~\bibnamefont{Xiao}},
  \bibinfo{author}{\bibfnamefont{M.}~\bibnamefont{Ke}},
  \bibinfo{author}{\bibfnamefont{F.}~\bibnamefont{Zhang}}, \bibnamefont{and}
  \bibinfo{author}{\bibfnamefont{Z.}~\bibnamefont{Liu}}, \bibinfo{journal}{Nat.
  Commun.} \textbf{\bibinfo{volume}{11}}, \bibinfo{pages}{1820}
  (\bibinfo{year}{2020}).

\bibitem[{\citenamefont{He et~al.}(2018)\citenamefont{He, Qiu, Ye, Cai, Fan,
  Ke, Zhang, and Liu}}]{49nat}
\bibinfo{author}{\bibfnamefont{H.}~\bibnamefont{He}},
  \bibinfo{author}{\bibfnamefont{C.}~\bibnamefont{Qiu}},
  \bibinfo{author}{\bibfnamefont{L.}~\bibnamefont{Ye}},
  \bibinfo{author}{\bibfnamefont{X.}~\bibnamefont{Cai}},
  \bibinfo{author}{\bibfnamefont{X.}~\bibnamefont{Fan}},
  \bibinfo{author}{\bibfnamefont{M.}~\bibnamefont{Ke}},
  \bibinfo{author}{\bibfnamefont{F.}~\bibnamefont{Zhang}}, \bibnamefont{and}
  \bibinfo{author}{\bibfnamefont{Z.}~\bibnamefont{Liu}},
  \bibinfo{journal}{Nature} \textbf{\bibinfo{volume}{560}}, \bibinfo{pages}{61}
  (\bibinfo{year}{2018}).

\bibitem[{\citenamefont{Shi et~al.}(2019)\citenamefont{Shi, Wieder, Meyerheim,
  Sun, Zhang, Li, Shen, Qi, Yang, Jena et~al.}}]{50}
\bibinfo{author}{\bibfnamefont{W.}~\bibnamefont{Shi}},
  \bibinfo{author}{\bibfnamefont{B.~J.} \bibnamefont{Wieder}},
  \bibinfo{author}{\bibfnamefont{H.}~\bibnamefont{Meyerheim}},
  \bibinfo{author}{\bibfnamefont{Y.}~\bibnamefont{Sun}},
  \bibinfo{author}{\bibfnamefont{Y.}~\bibnamefont{Zhang}},
  \bibinfo{author}{\bibfnamefont{Y.}~\bibnamefont{Li}},
  \bibinfo{author}{\bibfnamefont{L.}~\bibnamefont{Shen}},
  \bibinfo{author}{\bibfnamefont{Y.}~\bibnamefont{Qi}},
  \bibinfo{author}{\bibfnamefont{L.}~\bibnamefont{Yang}},
  \bibinfo{author}{\bibfnamefont{J.}~\bibnamefont{Jena}}, \bibnamefont{et~al.},
  \bibinfo{journal}{arXiv preprint arXiv:1909.04037}  (\bibinfo{year}{2019}).

\bibitem[{\citenamefont{Li et~al.}(2019)\citenamefont{Li, Deng, Fu, Li, Ma,
  Han, Zhou, Zhou, and Yao}}]{51}
\bibinfo{author}{\bibfnamefont{X.-P.} \bibnamefont{Li}},
  \bibinfo{author}{\bibfnamefont{K.}~\bibnamefont{Deng}},
  \bibinfo{author}{\bibfnamefont{B.}~\bibnamefont{Fu}},
  \bibinfo{author}{\bibfnamefont{Y.}~\bibnamefont{Li}},
  \bibinfo{author}{\bibfnamefont{D.}~\bibnamefont{Ma}},
  \bibinfo{author}{\bibfnamefont{J.}~\bibnamefont{Han}},
  \bibinfo{author}{\bibfnamefont{J.}~\bibnamefont{Zhou}},
  \bibinfo{author}{\bibfnamefont{S.}~\bibnamefont{Zhou}}, \bibnamefont{and}
  \bibinfo{author}{\bibfnamefont{Y.}~\bibnamefont{Yao}},
  \bibinfo{journal}{arXiv preprint arXiv:1909.12178}  (\bibinfo{year}{2019}).

\bibitem[{\citenamefont{Zhang et~al.}(2020)\citenamefont{Zhang, Takahashi,
  Fang, and Murakami}}]{52}
\bibinfo{author}{\bibfnamefont{T.}~\bibnamefont{Zhang}},
  \bibinfo{author}{\bibfnamefont{R.}~\bibnamefont{Takahashi}},
  \bibinfo{author}{\bibfnamefont{C.}~\bibnamefont{Fang}}, \bibnamefont{and}
  \bibinfo{author}{\bibfnamefont{S.}~\bibnamefont{Murakami}},
  \bibinfo{journal}{arXiv preprint arXiv:2004.02562}  (\bibinfo{year}{2020}).

\bibitem[{\citenamefont{Liu et~al.}(work in progress)}]{53}
\bibinfo{author}{\bibfnamefont{Q.-B.} \bibnamefont{Liu}} \bibnamefont{et~al.}
  (\bibinfo{year}{work in progress}).

\bibitem[{\citenamefont{Schindler et~al.}(2018)\citenamefont{Schindler, Cook,
  Vergniory, Wang, Parkin, Bernevig, and Neupert}}]{33}
\bibinfo{author}{\bibfnamefont{F.}~\bibnamefont{Schindler}},
  \bibinfo{author}{\bibfnamefont{A.~M.} \bibnamefont{Cook}},
  \bibinfo{author}{\bibfnamefont{M.~G.} \bibnamefont{Vergniory}},
  \bibinfo{author}{\bibfnamefont{Z.}~\bibnamefont{Wang}},
  \bibinfo{author}{\bibfnamefont{S.~S.} \bibnamefont{Parkin}},
  \bibinfo{author}{\bibfnamefont{B.~A.} \bibnamefont{Bernevig}},
  \bibnamefont{and} \bibinfo{author}{\bibfnamefont{T.}~\bibnamefont{Neupert}},
  \bibinfo{journal}{Science advances} \textbf{\bibinfo{volume}{4}},
  \bibinfo{pages}{eaat0346} (\bibinfo{year}{2018}).

\bibitem[{\citenamefont{Hautier
  et~al.}(2011{\natexlab{a}})\citenamefont{Hautier, Fischer, Ehrlacher, Jain,
  and Ceder}}]{34}
\bibinfo{author}{\bibfnamefont{G.}~\bibnamefont{Hautier}},
  \bibinfo{author}{\bibfnamefont{C.}~\bibnamefont{Fischer}},
  \bibinfo{author}{\bibfnamefont{V.}~\bibnamefont{Ehrlacher}},
  \bibinfo{author}{\bibfnamefont{A.}~\bibnamefont{Jain}}, \bibnamefont{and}
  \bibinfo{author}{\bibfnamefont{G.}~\bibnamefont{Ceder}},
  \bibinfo{journal}{Inorganic chemistry} \textbf{\bibinfo{volume}{50}},
  \bibinfo{pages}{656} (\bibinfo{year}{2011}{\natexlab{a}}).

\bibitem[{\citenamefont{Soluyanov and Vanderbilt}(2011)}]{46}
\bibinfo{author}{\bibfnamefont{A.~A.} \bibnamefont{Soluyanov}}
  \bibnamefont{and}
  \bibinfo{author}{\bibfnamefont{D.}~\bibnamefont{Vanderbilt}},
  \bibinfo{journal}{Physical Review B} \textbf{\bibinfo{volume}{83}},
  \bibinfo{pages}{235401} (\bibinfo{year}{2011}).

\bibitem[{\citenamefont{Yu et~al.}(2011)\citenamefont{Yu, Qi, Bernevig, Fang,
  and Dai}}]{47}
\bibinfo{author}{\bibfnamefont{R.}~\bibnamefont{Yu}},
  \bibinfo{author}{\bibfnamefont{X.~L.} \bibnamefont{Qi}},
  \bibinfo{author}{\bibfnamefont{A.}~\bibnamefont{Bernevig}},
  \bibinfo{author}{\bibfnamefont{Z.}~\bibnamefont{Fang}}, \bibnamefont{and}
  \bibinfo{author}{\bibfnamefont{X.}~\bibnamefont{Dai}},
  \bibinfo{journal}{Phys. Rev. B} \textbf{\bibinfo{volume}{84}},
  \bibinfo{pages}{075119} (\bibinfo{year}{2011}),
  \urlprefix\url{https://link.aps.org/doi/10.1103/PhysRevB.84.075119}.

\bibitem[{\citenamefont{Fang et~al.}(2016)\citenamefont{Fang, Lu, Liu, and
  Fu}}]{36}
\bibinfo{author}{\bibfnamefont{C.}~\bibnamefont{Fang}},
  \bibinfo{author}{\bibfnamefont{L.}~\bibnamefont{Lu}},
  \bibinfo{author}{\bibfnamefont{J.}~\bibnamefont{Liu}}, \bibnamefont{and}
  \bibinfo{author}{\bibfnamefont{L.}~\bibnamefont{Fu}},
  \bibinfo{journal}{Nature Physics} \textbf{\bibinfo{volume}{12}},
  \bibinfo{pages}{936} (\bibinfo{year}{2016}).

\bibitem[{\citenamefont{Kresse}(1995)}]{37}
\bibinfo{author}{\bibfnamefont{G.}~\bibnamefont{Kresse}},
  \bibinfo{journal}{Journal of Non-Crystalline Solids}
  \textbf{\bibinfo{volume}{192}}, \bibinfo{pages}{222} (\bibinfo{year}{1995}).

\bibitem[{\citenamefont{Kresse and Hafner}(1994)}]{38}
\bibinfo{author}{\bibfnamefont{G.}~\bibnamefont{Kresse}} \bibnamefont{and}
  \bibinfo{author}{\bibfnamefont{J.}~\bibnamefont{Hafner}},
  \bibinfo{journal}{Physical Review B} \textbf{\bibinfo{volume}{49}},
  \bibinfo{pages}{14251} (\bibinfo{year}{1994}).

\bibitem[{\citenamefont{Kresse and Furthm{\"u}ller}(1996)}]{39}
\bibinfo{author}{\bibfnamefont{G.}~\bibnamefont{Kresse}} \bibnamefont{and}
  \bibinfo{author}{\bibfnamefont{J.}~\bibnamefont{Furthm{\"u}ller}},
  \bibinfo{journal}{Physical review B} \textbf{\bibinfo{volume}{54}},
  \bibinfo{pages}{11169} (\bibinfo{year}{1996}).

\bibitem[{\citenamefont{Kohn and Sham}(1965)}]{40}
\bibinfo{author}{\bibfnamefont{W.}~\bibnamefont{Kohn}} \bibnamefont{and}
  \bibinfo{author}{\bibfnamefont{L.~J.} \bibnamefont{Sham}},
  \bibinfo{journal}{Physical review} \textbf{\bibinfo{volume}{140}},
  \bibinfo{pages}{A1133} (\bibinfo{year}{1965}).

\bibitem[{\citenamefont{Bl{\"o}chl}(1994)}]{41}
\bibinfo{author}{\bibfnamefont{P.~E.} \bibnamefont{Bl{\"o}chl}},
  \bibinfo{journal}{Physical review B} \textbf{\bibinfo{volume}{50}},
  \bibinfo{pages}{17953} (\bibinfo{year}{1994}).

\bibitem[{\citenamefont{Perdew et~al.}(1996)\citenamefont{Perdew, Burke, and
  Ernzerhof}}]{42}
\bibinfo{author}{\bibfnamefont{J.~P.} \bibnamefont{Perdew}},
  \bibinfo{author}{\bibfnamefont{K.}~\bibnamefont{Burke}}, \bibnamefont{and}
  \bibinfo{author}{\bibfnamefont{M.}~\bibnamefont{Ernzerhof}},
  \bibinfo{journal}{Physical review letters} \textbf{\bibinfo{volume}{77}},
  \bibinfo{pages}{3865} (\bibinfo{year}{1996}).

\bibitem[{\citenamefont{Togo and Tanaka}(2015)}]{43}
\bibinfo{author}{\bibfnamefont{A.}~\bibnamefont{Togo}} \bibnamefont{and}
  \bibinfo{author}{\bibfnamefont{I.}~\bibnamefont{Tanaka}},
  \bibinfo{journal}{Scripta Materialia} \textbf{\bibinfo{volume}{108}},
  \bibinfo{pages}{1} (\bibinfo{year}{2015}).

\bibitem[{\citenamefont{Wu et~al.}(2018)\citenamefont{Wu, Zhang, Song, Troyer,
  and Soluyanov}}]{44}
\bibinfo{author}{\bibfnamefont{Q.}~\bibnamefont{Wu}},
  \bibinfo{author}{\bibfnamefont{S.}~\bibnamefont{Zhang}},
  \bibinfo{author}{\bibfnamefont{H.-F.} \bibnamefont{Song}},
  \bibinfo{author}{\bibfnamefont{M.}~\bibnamefont{Troyer}}, \bibnamefont{and}
  \bibinfo{author}{\bibfnamefont{A.~A.} \bibnamefont{Soluyanov}},
  \bibinfo{journal}{Computer Physics Communications}
  \textbf{\bibinfo{volume}{224}}, \bibinfo{pages}{405} (\bibinfo{year}{2018}).

\bibitem[{\citenamefont{Sancho et~al.}(1985)\citenamefont{Sancho, Sancho,
  Sancho, and Rubio}}]{45}
\bibinfo{author}{\bibfnamefont{M.~L.} \bibnamefont{Sancho}},
  \bibinfo{author}{\bibfnamefont{J.~L.} \bibnamefont{Sancho}},
  \bibinfo{author}{\bibfnamefont{J.~L.} \bibnamefont{Sancho}},
  \bibnamefont{and} \bibinfo{author}{\bibfnamefont{J.}~\bibnamefont{Rubio}},
  \bibinfo{journal}{Journal of Physics F: Metal Physics}
  \textbf{\bibinfo{volume}{15}}, \bibinfo{pages}{851} (\bibinfo{year}{1985}).

\bibitem[{\citenamefont{Gao et~al.}(2020)\citenamefont{Gao, Wu, Persson, and
  Wang}}]{gao2020irvsp}
\bibinfo{author}{\bibfnamefont{J.}~\bibnamefont{Gao}},
  \bibinfo{author}{\bibfnamefont{Q.}~\bibnamefont{Wu}},
  \bibinfo{author}{\bibfnamefont{C.}~\bibnamefont{Persson}}, \bibnamefont{and}
  \bibinfo{author}{\bibfnamefont{Z.}~\bibnamefont{Wang}},
  \bibinfo{journal}{arXiv preprint arXiv:2002.04032}  (\bibinfo{year}{2020}).

\bibitem[{\citenamefont{Hautier
  et~al.}(2011{\natexlab{b}})\citenamefont{Hautier, Fischer, Ehrlacher, Jain,
  and Ceder}}]{sf1}
\bibinfo{author}{\bibfnamefont{G.}~\bibnamefont{Hautier}},
  \bibinfo{author}{\bibfnamefont{C.}~\bibnamefont{Fischer}},
  \bibinfo{author}{\bibfnamefont{V.}~\bibnamefont{Ehrlacher}},
  \bibinfo{author}{\bibfnamefont{A.}~\bibnamefont{Jain}}, \bibnamefont{and}
  \bibinfo{author}{\bibfnamefont{G.}~\bibnamefont{Ceder}},
  \bibinfo{journal}{Inorganic chemistry} \textbf{\bibinfo{volume}{50}},
  \bibinfo{pages}{656} (\bibinfo{year}{2011}{\natexlab{b}}).

\bibitem[{\citenamefont{Bortz and Jansen}(1993)}]{sf2}
\bibinfo{author}{\bibfnamefont{M.}~\bibnamefont{Bortz}} \bibnamefont{and}
  \bibinfo{author}{\bibfnamefont{M.}~\bibnamefont{Jansen}},
  \bibinfo{journal}{Zeitschrift f{\"u}r anorganische und allgemeine Chemie}
  \textbf{\bibinfo{volume}{619}}, \bibinfo{pages}{1446} (\bibinfo{year}{1993}).

\bibitem[{\citenamefont{Martens and Hoppe}(1977)}]{sf3}
\bibinfo{author}{\bibfnamefont{K.-P.} \bibnamefont{Martens}} \bibnamefont{and}
  \bibinfo{author}{\bibfnamefont{R.}~\bibnamefont{Hoppe}},
  \bibinfo{journal}{Zeitschrift f{\"u}r anorganische und allgemeine Chemie}
  \textbf{\bibinfo{volume}{437}}, \bibinfo{pages}{116} (\bibinfo{year}{1977}).

\bibitem[{\citenamefont{Niu et~al.}(2014)\citenamefont{Niu, Wang, and
  Wang}}]{sf4}
\bibinfo{author}{\bibfnamefont{C.-Y.} \bibnamefont{Niu}},
  \bibinfo{author}{\bibfnamefont{X.-Q.} \bibnamefont{Wang}}, \bibnamefont{and}
  \bibinfo{author}{\bibfnamefont{J.-T.} \bibnamefont{Wang}},
  \bibinfo{journal}{The Journal of chemical physics}
  \textbf{\bibinfo{volume}{140}}, \bibinfo{pages}{054514}
  (\bibinfo{year}{2014}).

\end{thebibliography}

%
%
%

\ \\
\newpage

\ \\
\newpage

\begin{widetext}
\beginsupplement{}
\section*{\Large{Supplementary Materials for ``Symmetry-Enforced Weyl Phonons"}}

\subsection*{A. A complete list of symmetry-protected Weyl points}
In the main text, we are focused on some high-symmetry $k$-points, where only the two-dimensional irrep of Weyl points (WPs) is allowed.
In this section, we have presented a complete list of high-symmetry $k$-points in Table~\ref{tab:SPWP}, wherethe two-dimensional irrep of WPs can be protected in bosonic systems. We call them the symmetry-protected WPs. We find that these two-fold WPs on the list can host Chern numbers of 
$\pm 1$ [\ie (monopole) WPs], $\pm 2$ [\ie double WPs] and $\pm 4$ [\ie quadruple WPs], respectively. 
So far, the quadruple WPs with Chern numbers of $\pm 4$ (\ie $|C|=4$) have been rarely reported\cite{52,53}. Guided by the list in Table~\ref{tab:SPWP}, the quadruple WPs can be checked in future experiments.

\begin{table}[!h]
\caption{A complete list of symmetry-protected two-fold Weyl points in bosonic systems in the presence of time-reversal symmetry. The first and the second columns indicate the space group (SG) number and the corresponding high-symmetry $k$-point [\eg $\Gamma$(GM), Z, A, M, H, K, $etc.$], respectively. The third and fourth columns show the abstract group (AG), which the little group of the $k$-point is isomorphic with, and the corresponding irreps. The fifth column indicates the Chern numbers ($C$) of these two-fold Weyl points.
See more details for the AGs and their character tables in Ref.~\cite{28}.
The asterisks indicate the high-symmetry $k$-points, where \emph{only} the two-dimensional irrep of Weyl nodes is allowed.
}\label{tab:SPWP}
\begin{tabular}{c|c|c|c|cp{0.5cm}c|c|c|c|cp{0.5cm}c|c|c|c|cp{0.5cm}c|c|c|c|c}
\hline
\hline
   SG &  $k$ & AG & irreps & $C$
&& SG &  $k$ & AG & irreps & $C$
&& SG &  $k$ & AG & irreps & $C$
&& SG &  $k$ & AG & irreps & $C$\\
\cline{1-5}
\cline{7-11}
\cline{13-17}
\cline{19-23}
\cline{1-5}
\cline{7-11}
\cline{13-17}
\cline{19-23}
    * 24&  W  &   $G_{16}^ 7$   &$ 9   $ & $\pm1$
 &&   75& GM  &   $G_{ 4}^ 1$   & $2+ 4$ & $\pm2$
 &&   75&  M  &   $G_{ 4}^ 1$   & $2+ 4$ & $\pm2$
 &&   75&  Z  &   $G_{ 4}^ 1$   & $2+ 4$ & $\pm2$
 \\   75&  A  &   $G_{ 4}^ 1$   & $2+ 4$ & $\pm2$
 &&   76& GM  &   $G_{ 4}^ 1$   & $2+ 4$ & $\pm2$
 &&   76&  M  &   $G_{ 4}^ 1$   & $2+ 4$ & $\pm2$
 &&   77& GM  &   $G_{ 4}^ 1$   & $2+ 4$ & $\pm2$
 \\   77&  M  &   $G_{ 4}^ 1$   & $2+ 4$ & $\pm2$
 &&   77&  Z  &   $G_{ 4}^ 1$   & $2+ 4$ & $\pm2$
 &&   77&  A  &   $G_{ 4}^ 1$   & $2+ 4$ & $\pm2$
 &&   78& GM  &   $G_{ 4}^ 1$   & $2+ 4$ & $\pm2$
 \\   78&  M  &   $G_{ 4}^ 1$   & $2+ 4$ & $\pm2$
 &&   79& GM  &   $G_{ 4}^ 1$   & $2+ 4$ & $\pm2$
 &&   79&  Z  &   $G_{ 4}^ 1$   & $2+ 4$ & $\pm2$
 &&   79&  P  &   $G_{ 2}^ 1$   & $2+ 2$ & $\pm2$
 \\   80& GM  &   $G_{ 4}^ 1$   & $2+ 4$ & $\pm2$
 &&   80&  Z  &   $G_{ 4}^ 1$   & $2+ 4$ & $\pm2$
 && * 80&  P  &   $G_{ 2}^ 1$   &$ 1+ 2$ & $\pm1$ 
 &&   89& GM  &   $G_{ 8}^ 4$   & $5   $ & $\pm2$
 \\   89&  M  &   $G_{ 8}^ 4$   & $5   $ & $\pm2$
 &&   89&  Z  &   $G_{ 8}^ 4$   & $5   $ & $\pm2$
 &&   89&  A  &   $G_{ 8}^ 4$   & $5   $ & $\pm2$
 &&   90& GM  &   $G_{ 8}^ 4$   & $5   $ & $\pm2$
 \\   90&  Z  &   $G_{ 8}^ 4$   & $5   $ & $\pm2$
 &&   91& GM  &   $G_{ 8}^ 4$   & $5   $ & $\pm2$
 &&   91&  M  &   $G_{ 8}^ 4$   & $5   $ & $\pm2$
 &&   92& GM  &   $G_{ 8}^ 4$   & $5   $ & $\pm2$
 \\   93& GM  &   $G_{ 8}^ 4$   & $5   $ & $\pm2$
 &&   93&  M  &   $G_{ 8}^ 4$   & $5   $ & $\pm2$
 &&   93&  Z  &   $G_{16}^ 9$   & $5   $ & $\pm2$
 &&   93&  A  &   $G_{16}^ 9$   & $5   $ & $\pm2$
 \\   94& GM  &   $G_{ 8}^ 4$   & $5   $ & $\pm2$
 &&   94&  Z  &   $G_{16}^ 9$   & $5   $ & $\pm2$
 &&   95& GM  &   $G_{ 8}^ 4$   & $5   $ & $\pm2$
 &&   95&  M  &   $G_{ 8}^ 4$   & $5   $ & $\pm2$
 \\   96& GM  &   $G_{ 8}^ 4$   & $5   $ & $\pm2$
 &&   97& GM  &   $G_{ 8}^ 4$   & $5   $ & $\pm2$
 &&   97&  Z  &   $G_{ 8}^ 4$   & $5   $ & $\pm2$
 &&   97&  P  &   $G_{ 4}^ 2$   & $3+ 4$ & $\pm2$  
 \\   98& GM  &   $G_{ 8}^ 4$   & $5   $ & $\pm2$
 &&   98&  Z  &   $G_{ 8}^ 4$   & $5   $ & $\pm2$
 && * 98&  P  &   $G_{16}^ 7$   &$10   $ & $\pm1$  
 &&  143& GM  &   $G_{ 3}^ 1$   & $2+ 3$ & $\pm2$
 \\  143&  A  &   $G_{ 3}^ 1$   & $2+ 3$ & $\pm2$
 &&  144& GM  &   $G_{ 3}^ 1$   & $2+ 3$ & $\pm2$
 &&  144&  A  &   $G_{ 3}^ 1$   & $2+ 3$ & $\pm2$
 &&  145& GM  &   $G_{ 3}^ 1$   & $2+ 3$ & $\pm2$
 \\  145&  A  &   $G_{ 3}^ 1$   & $2+ 3$ & $\pm2$
 &&  146& GM  &   $G_{ 3}^ 1$   & $2+ 3$ & $\pm2$
 &&  146&  Z  &   $G_{ 3}^ 1$   & $2+ 3$ & $\pm2$
 &&  149& GM  &   $G_{ 6}^ 2$   & $3   $ & $\pm2$
 \\  149&  A  &   $G_{ 6}^ 2$   & $3   $ & $\pm2$
 &&  150& GM  &   $G_{ 6}^ 2$   & $3   $ & $\pm2$
 &&  150&  A  &   $G_{ 6}^ 2$   & $3   $ & $\pm2$
 &&  150&  K  &   $G_{ 6}^ 2$   & $3   $ & $\pm1$
 \\  150&  H  &   $G_{ 6}^ 2$   & $3   $ & $\pm1$
 &&  151& GM  &   $G_{ 6}^ 2$   & $3   $ & $\pm2$
 &&  151&  A  &   $G_{ 6}^ 2$   & $3   $ & $\pm2$
 &&  152& GM  &   $G_{ 6}^ 2$   & $3   $ & $\pm2$
 \\  152&  A  &   $G_{ 6}^ 2$   & $3   $ & $\pm2$
 &&  152&  K  &   $G_{ 6}^ 2$   & $3   $ & $\pm1$
 &&  152&  H  &   $G_{ 6}^ 2$   & $3   $ & $\pm1$
 &&  153& GM  &   $G_{ 6}^ 2$   & $3   $ & $\pm2$
 \\  153&  A  &   $G_{ 6}^ 2$   & $3   $ & $\pm2$
 &&  154& GM  &   $G_{ 6}^ 2$   & $3   $ & $\pm2$
 &&  154&  A  &   $G_{ 6}^ 2$   & $3   $ & $\pm2$
 &&  154&  K  &   $G_{ 6}^ 2$   & $3   $ & $\pm1$
 \\  154&  H  &   $G_{ 6}^ 2$   & $3   $ & $\pm1$
 &&  155& GM  &   $G_{ 6}^ 2$   & $3   $ & $\pm2$
 &&  155&  Z  &   $G_{ 6}^ 2$   & $3   $ & $\pm2$
 &&  168& GM  &   $G_{ 6}^ 1$   & $2+ 6$ & $\pm2$
 \\  168& GM  &   $G_{ 6}^ 1$   & $3+ 5$ & $\pm2$
 &&  168&  A  &   $G_{ 6}^ 1$   & $2+ 6$ & $\pm2$
 &&  168&  A  &   $G_{ 6}^ 1$   & $3+ 5$ & $\pm2$
 &&  168&  K  &   $G_{ 3}^ 1$   & $2+ 3$ & $\pm1$
 \\  168&  H  &   $G_{ 3}^ 1$   & $2+ 3$ & $\pm1$
 &&  169& GM  &   $G_{ 6}^ 1$   & $2+ 6$ & $\pm2$
 &&  169& GM  &   $G_{ 6}^ 1$   & $3+ 5$ & $\pm2$
 &&  169&  K  &   $G_{ 3}^ 1$   & $2+ 3$ & $\pm1$
 \\  170& GM  &   $G_{ 6}^ 1$   & $2+ 6$ & $\pm2$
 &&  170& GM  &   $G_{ 6}^ 1$   & $3+ 5$ & $\pm2$
 &&  170&  K  &   $G_{ 3}^ 1$   & $2+ 3$ & $\pm1$
 &&  171& GM  &   $G_{ 6}^ 1$   & $2+ 6$ & $\pm2$
 \\  171& GM  &   $G_{ 6}^ 1$   & $3+ 5$ & $\pm2$
 &&  171&  A  &   $G_{ 6}^ 1$   & $2+ 6$ & $\pm2$
 &&  171&  A  &   $G_{ 6}^ 1$   & $3+ 5$ & $\pm2$
 &&  171&  K  &   $G_{ 3}^ 1$   & $2+ 3$ & $\pm1$
 \\  171&  H  &   $G_{ 3}^ 1$   & $2+ 3$ & $\pm1$
 &&  172& GM  &   $G_{ 6}^ 1$   & $2+ 6$ & $\pm2$
 &&  172& GM  &   $G_{ 6}^ 1$   & $3+ 5$ & $\pm2$
 &&  172&  A  &   $G_{ 6}^ 1$   & $2+ 6$ & $\pm2$
 \\  172&  A  &   $G_{ 6}^ 1$   & $3+ 5$ & $\pm2$
 &&  172&  K  &   $G_{ 3}^ 1$   & $2+ 3$ & $\pm1$
 &&  172&  H  &   $G_{ 3}^ 1$   & $2+ 3$ & $\pm1$
 &&  173& GM  &   $G_{ 6}^ 1$   & $2+ 6$ & $\pm2$
 \\  173& GM  &   $G_{ 6}^ 1$   & $3+ 5$ & $\pm2$
 &&  173&  K  &   $G_{ 3}^ 1$   & $2+ 3$ & $\pm1$  
 &&  177& GM  &   $G_{12}^ 3$   & $5   $ & $\pm2$
 &&  177& GM  &   $G_{12}^ 3$   & $6   $ & $\pm2$
 \\  177&  A  &   $G_{12}^ 3$   & $5   $ & $\pm2$
 &&  177&  A  &   $G_{12}^ 3$   & $6   $ & $\pm2$
 &&  177&  K  &   $G_{ 6}^ 2$   & $3   $ & $\pm1$
 &&  177&  H  &   $G_{ 6}^ 2$   & $3   $ & $\pm1$
 \\  178& GM  &   $G_{12}^ 3$   & $5   $ & $\pm2$
 &&  178& GM  &   $G_{12}^ 3$   & $6   $ & $\pm2$
 &&  178&  K  &   $G_{ 6}^ 2$   & $3   $ & $\pm1$
 &&  179& GM  &   $G_{12}^ 3$   & $5   $ & $\pm2$
 \\  179& GM  &   $G_{12}^ 3$   & $6   $ & $\pm2$
 &&  179&  K  &   $G_{ 6}^ 2$   & $3   $ & $\pm1$
 &&  180& GM  &   $G_{12}^ 3$   & $5   $ & $\pm2$
 &&  180& GM  &   $G_{12}^ 3$   & $6   $ & $\pm2$
 \\  180&  A  &   $G_{12}^ 3$   & $5   $ & $\pm2$
 &&  180&  A  &   $G_{12}^ 3$   & $6   $ & $\pm2$
 &&  180&  K  &   $G_{ 6}^ 2$   & $3   $ & $\pm1$
 &&  180&  H  &   $G_{ 6}^ 2$   & $3   $ & $\pm1$
 \\  181& GM  &   $G_{12}^ 3$   & $5   $ & $\pm2$
 &&  181& GM  &   $G_{12}^ 3$   & $6   $ & $\pm2$
 &&  181&  A  &   $G_{12}^ 3$   & $5   $ & $\pm2$
 &&  181&  A  &   $G_{12}^ 3$   & $6   $ & $\pm2$
 \\  181&  K  &   $G_{ 6}^ 2$   & $3   $ & $\pm1$
 &&  181&  H  &   $G_{ 6}^ 2$   & $3   $ & $\pm1$
 &&  182& GM  &   $G_{12}^ 3$   & $5   $ & $\pm2$
 &&  182& GM  &   $G_{12}^ 3$   & $6   $ & $\pm2$
 \\  182&  K  &   $G_{ 6}^ 2$   & $3   $ & $\pm1$
 &&  195& GM  &   $G_{12}^ 5$   & $2+ 3$ & $\pm4$
 &&  195&  R  &   $G_{12}^ 5$   & $2+ 3$ & $\pm4$
 &&  196& GM  &   $G_{12}^ 5$   & $2+ 3$ & $\pm4$
 \\  196&  L  &   $G_{ 3}^ 1$   & $2+ 3$ & $\pm2$
 &&  197& GM  &   $G_{12}^ 5$   & $2+ 3$ & $\pm4$
 &&  197&  H  &   $G_{12}^ 5$   & $2+ 3$ & $\pm4$
 &&  198& GM  &   $G_{12}^ 5$   & $2+ 3$ & $\pm4$
 \\  199& GM  &   $G_{12}^ 5$   & $2+ 3$ & $\pm4$
 &&  199&  H  &   $G_{24}^ 8$   & $5+ 6$ & $\pm4$
 && *199&  P  &   $G_{48}^ 3$   &$ 7   $ & $\pm1$
 && *199&  P  &   $G_{48}^ 3$   &$ 8   $ & $\pm1$
 \\ *199&  P  &   $G_{48}^ 3$   &$ 9   $ & $\pm1$
 &&  207& GM  &   $G_{24}^ 7$   & $3   $ & $\pm4$
 &&  207&  X  &   $G_{ 8}^ 4$   & $5   $ & $\pm2$
 &&  207&  M  &   $G_{ 8}^ 4$   & $5   $ & $\pm2$
 \\  207&  R  &   $G_{24}^ 7$   & $3   $ & $\pm4$
 &&  208& GM  &   $G_{24}^ 7$   & $3   $ & $\pm4$
 &&  208&  X  &   $G_{ 8}^ 4$   & $5   $ & $\pm2$
 &&  208&  M  &   $G_{ 8}^ 4$   & $5   $ & $\pm2$
 \\  208&  R  &   $G_{24}^ 7$   & $3   $ & $\pm4$
 &&  209& GM  &   $G_{24}^ 7$   & $3   $ & $\pm4$
 &&  209&  X  &   $G_{ 8}^ 4$   & $5   $ & $\pm2$
 &&  209&  L  &   $G_{ 6}^ 2$   & $3   $ & $\pm2$
 \\  209&  W  &   $G_{ 4}^ 2$   & $3+ 4$ & $\pm2$  
 &&  210& GM  &   $G_{24}^ 7$   & $3   $ & $\pm4$
 &&  210&  X  &   $G_{ 8}^ 4$   & $5   $ & $\pm2$
 &&  210&  L  &   $G_{ 6}^ 2$   & $3   $ & $\pm2$
 \\ *210&  W  &   $G_{16}^ 7$   &$10   $ & $\pm1$
 &&  211& GM  &   $G_{24}^ 7$   & $3   $ & $\pm4$
 &&  211&  H  &   $G_{24}^ 7$   & $3   $ & $\pm4$  
 &&  212& GM  &   $G_{24}^ 7$   & $3   $ & $\pm4$
 \\  213& GM  &   $G_{24}^ 7$   & $3   $ & $\pm4$
 &&  214& GM  &   $G_{24}^ 7$   & $3   $ & $\pm4$
 &&  214&  H  &   $G_{48}^ 7$   & $6   $ & $\pm4$
 && *214&  P  &   $G_{48}^ 3$   &$ 7   $ & $\pm1$
 \\ *214&  P  &   $G_{48}^ 3$   &$ 8   $ & $\pm1$
 && *214&  P  &   $G_{48}^ 3$   &$ 9   $ & $\pm1$  
\\
\hline
\hline
\end{tabular}
\end{table}


\subsection*{B. The phonon dispersions of K$_2$Sn$_2$O$_3$ with LOTO}
To illustrate that the three-fold spin-1 Weyl phonon at the high-symmetry point H is robust against the LOTO modification, the phononic dispersions of the material example K$_2$Sn$_2$O$_3$ in the presence of the polariton of LOTO splitting are also calculated and shown in Fig.~\ref{fig:s1}. The three bands including 39$^{th}$, 40$^{th}$ and 41$^{st}$ bands, which form the three-fold spin-1 Weyl phonon with LOTO, are also highlighted in the right panel of Fig.~\ref{fig:s1}. It is clearly seen that the polariton of LOTO splitting does {\emph not} split the three-fold degeneracy at H, supporting that the three-fold spin-1 Weyl phonon in the present material candidate can be observed in future experiments.

\begin{figure}[!htb]
\centering
\includegraphics[width=15.5 cm]{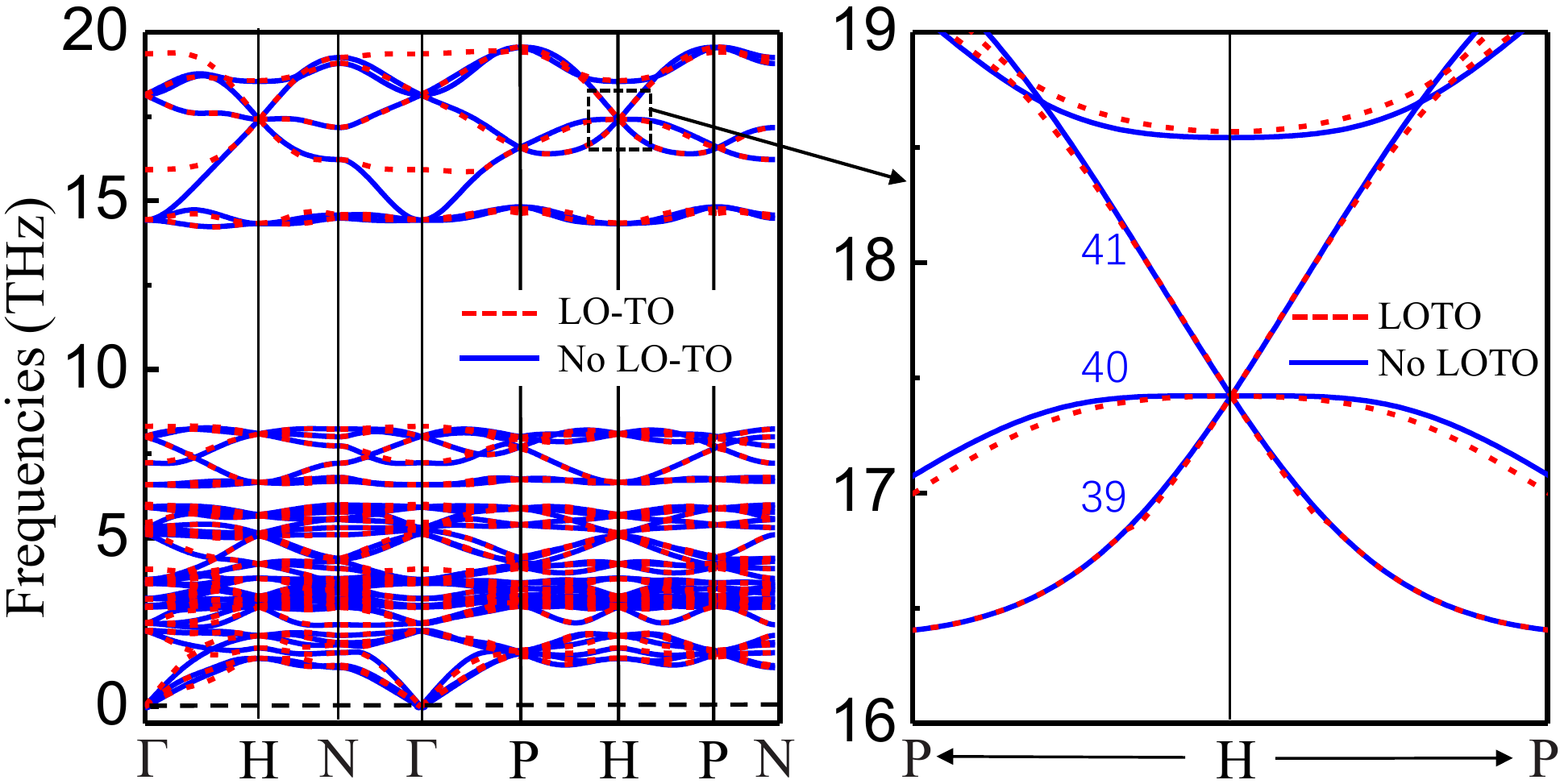}
\caption{The phonon dispersions of the crystal K$_2$Sn$_2$O$_3$ with and without the polariton of LOTO splitting are drawn in the left panel, and the 39$^{th}$, 40$^{th}$ and 41$^{st}$ bands around the high-symmetry point H are highlighted in the right panel.} \label{fig:s1}
\end{figure}

\clearpage

\subsection*{C. The degeneracies for the 40$^{th}$ and 41$^{st}$ phonon bands of K$_2$Sn$_2$O$_3$}

Between the 40$^{th}$ and 41$^{st}$ phonon bands of K$_2$Sn$_2$O$_3$ as shown in Fig.~\ref{fig:2}c in the main text, there are only two spin-1 Weyl phonons at the H and $\Gamma$ points, respectively. The Chern number of the spin-1 Weyl phonon at $\Gamma$ is $-2$. To show the nontrivial surface arc states associated with them, the isofrequency surface contours at 18.145 and 18.065 THz on the (110) surface are obtained and presented in Figs.~\ref{fig:s2}b and ~\ref{fig:s2}c, respectively. The surface arc states and the double helicoid surface states are clearly seen.

\begin{figure}[!htb]
\centering
\includegraphics[width=15.5 cm]{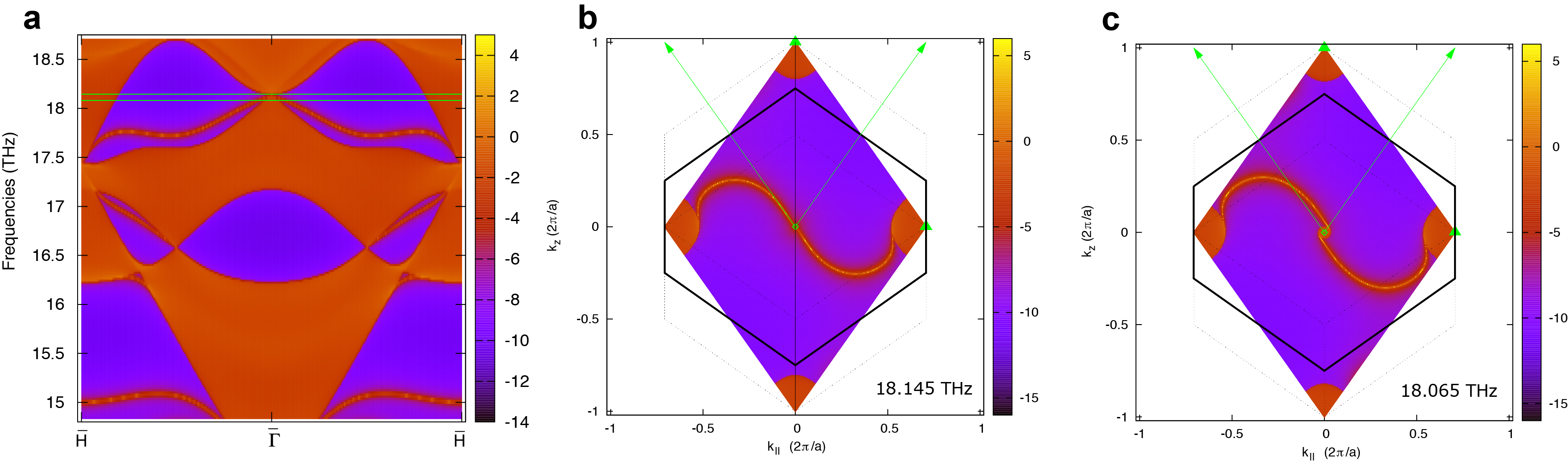}
\caption{\textbf{a,} The phonon dispersions on the (110) surface BZ of K$_2$Sn$_2$O$_3$. \textbf{b, c,} The surface arcs for the frequencies of 18.145 and 18.065 THz, respectively. One can see that the very long surface arcs and the double helicoid surface states appear clearly.} \label{fig:s2}
\end{figure}


\subsection*{D. The SEWPs at other space groups}

In this section, we present some other realistic material candidates in the SGs in Table~\ref{tab:SEWP} in the main text, to illustrate the existence of the SEWPs in realistic materials. Here we investigate the material examples PNO in SG 24, Ag$_3$BiO$_3$ in SG 80, LiAuO$_2$ in SG 98, K$_2$Pb$_2$O$_3$ in SG 199, SiO$_2$ in SG 210 and K$_6$ carbon in SG 214.

In the following subsections, we first draw their primitive unit cells and the first BZs to show the crystal structures, and then by using symmetry analysis and effective $k\cdot p$ models, we verify the guiding principle of the search of SEWPs as described in the main text. Finally, we calculated their phonon dispersions and DOS, and the Chern numbers of some nontrivial phonon bands to describe the existence of the WPs. The calculation methods are the same as those described in the main text.


\subsubsection*{\normalsize{1. The SEWPs in the material PNO in SG 24}}

The crystallographic data of PNO are adopted from Ref.~\cite{sf1} and the primitive cell is shown in Fig.~\ref{fig:s3}a, where the purple, gray and red atoms stand for P, N and O atoms, respectively. Each primitive cell contains 6 atoms with two P, two N and two O atoms, and the corresponding BZ is shown in Fig.~\ref{fig:s3}b. PNO belongs to the body-centered orthogonal structure with the SG I2$_1$2$_1$2$_1$ (No. 24).

SG 24 hosts only Weyl phonons at the W point (the high-symmetry points are defined in Ref.~\cite{28}, and it hosts one irreps of the AG $G_{16}^7$ in Table~\ref{tab:SEWP}. This SG has a body-centered orthogonal Bravais lattice. The operators $C_{2y}$ and $C_{2z}$ acting on the primitive lattice vectors ($\bt_1,\bt_2,\bt_3$) are presented~\cite{28} as follows:
\begin{equation*}
 \begin{split}
& C_{2z} \left(\begin{array}{ccc}
\bt_1 &\bt_2 &\bt_3
\end{array}\right)=
\left(\begin{array}
{ccc}
\bt_1 &\bt_2 &\bt_3
\end{array}\right)
\left[\begin{array}{ccc}
0 & 1 & -1 \\
1 & 0 & -1 \\
0 & 0 & -1 \end{array}\right],~
 C_{2y} \left(\begin{array}{ccc}
\bt_1 &\bt_2 &\bt_3
\end{array}\right)=
\left(\begin{array}
{ccc}
\bt_1 &\bt_2 &\bt_3
\end{array}\right)
\left[\begin{array}{ccc}
-1 & 0 & 0 \\
-1 & 0 & 1 \\
-1 & 1 & 0 \end{array}\right],
 \end{split}
\end{equation*}
Thus,
\begin{eqnarray*}
AB&=& \{C_{2z}|\frac{1}{2}0\frac{1}{2}\}\{C_{2y}|\frac{1}{2}\frac{1}{2}0\}= \{C_{2x}|1,\frac{1}{2},\frac{1}{2}\}\\
BA&=& \{C_{2y}|\frac{1}{2}\frac{1}{2}0\}\{C_{2z}|\frac{1}{2}0\frac{1}{2}\}=\{C_{2x}|0,\frac{1}{2},-\frac{1}{2}\} = \{E|-1,0,-1\} AB
\end{eqnarray*}
At the W point ($\frac{1}{4},\frac{1}{4},\frac{1}{4}$), the pure translation operator $\{E|-1,0,-1\}$ is expressed as $e^{2i\pi(-1+0-1)/4}=-1$. Therefore, we get $\{A,B\}=0$, which yields that all the phonon bands are doubly degenerate at W.

Next,we analyze a two-band model at the W point in SG 24. We have $A^2=\{E|000\}=1$, $B^2=\{E|000\}=1$ with $E$ identity operator and $\{A,B\}=0$. With the matrix representations of $A=\sigma_z$ and $B=\sigma_y$, the $k\cdot p$-invariant Hamiltonian is derived as (to the first order),
\[
H^{24}_P(\bk)=v_1\sigma_xk_x+v_2\sigma_yk_y+v_3\sigma_zk_z 
\]
with $\sigma_{x,y,z}$ Pauli matrices, $k_{x,y,z}$ momentum offset from the P point, and $v_{1,2,3}$ real coefficients. Obviously, it's a Weyl Hamiltonian. The P point of SG 98 and the W point of SG 210 in Table~\ref{tab:SEWP} with anti-unitary commutation relations share the similar results.

To confirm the above results from the symmetry analysis, the phononic dispersions and DOS of PNO from \emph{ab initio} calculations are illustrated in Fig.~\ref{fig:s3}c. One can see that two-fold Weyl phonons are localized at the high-symmetry point W, one of the corners of the first BZ. Note that the two linear crossing bands possess Chern numbers of $\pm$1, indicating its topologically nontrivial feature.

\begin{figure}[!hb]
\centering
\includegraphics[width=11.0 cm]{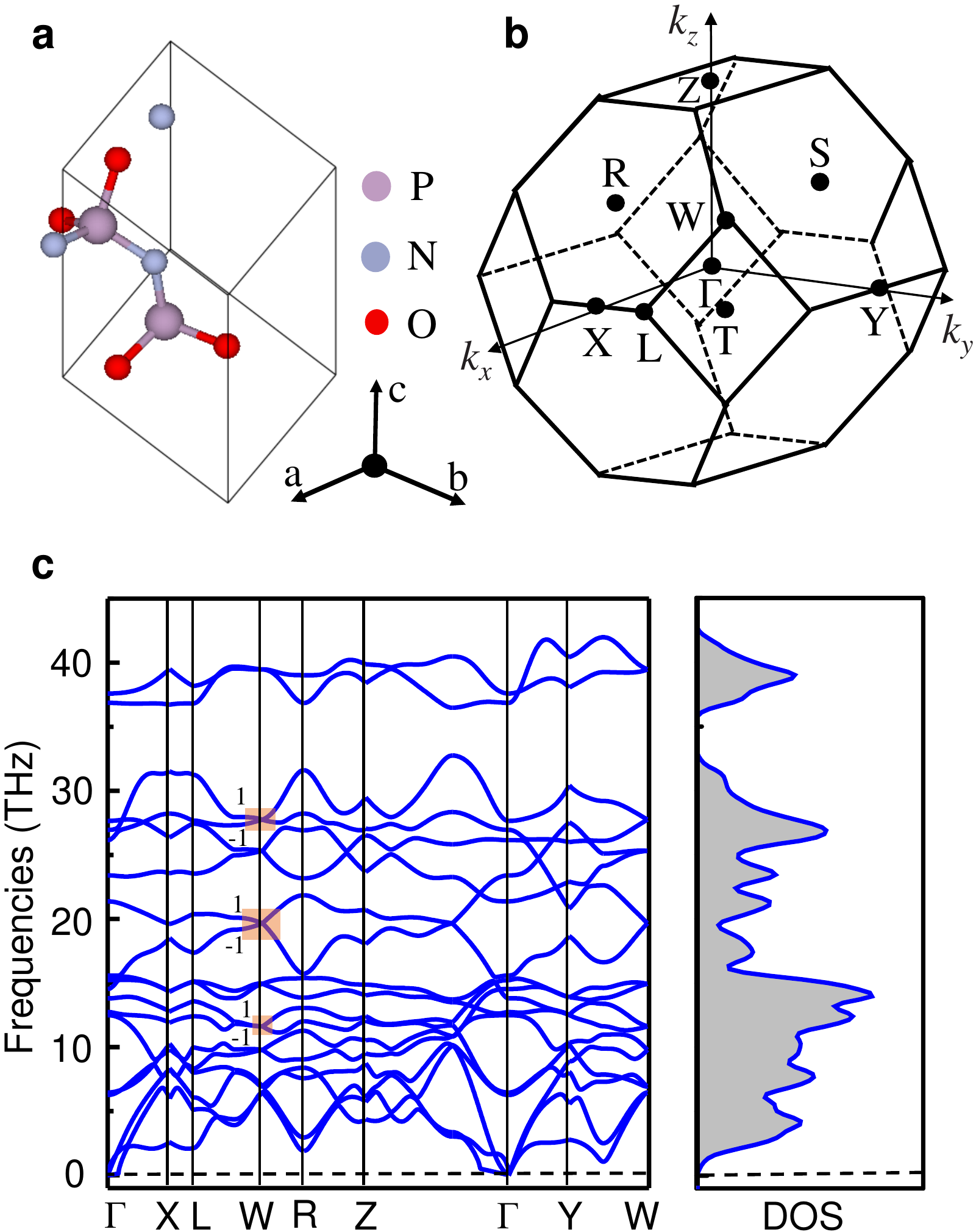}
\caption{A realistic candidate PNO in SG 24. \textbf{a,} A unit cell contains 2 \emph{P}, 2 \emph{N} and 2 O atoms. \textbf{b,} The first BZ of PNO. \textbf{c,} The phonon dispersions of PNO along high-symmetry directions. Some nontrivial bands have Chern numbers of $\pm$1 around W point, which are highlighted by orange color. It is clearly seen that the two-fold Weyl phonons are localized at the high-symmetry point W, indicating that the SEWPs exist in this material.} \label{fig:s3}
\end{figure}

\clearpage

\subsubsection*{\normalsize{2. The SEWPs in the material Ag$_3$BiO$_3$ in SG 80}}

The crystallographic data of Ag$_3$BiO$_3$ are adopted from Ref.~\cite{sf1} and the primitive cell is shown in Fig.~\ref{fig:s4}a, where the gray (rose red and red) atoms stand for Ag (Bi and O) atoms. Ag$_3$BiO$_3$ belongs to the body-centered tetragonal structure with SG I4$_1$ (No. 80)~\cite{sf2}. A primitive cell contains 56 atoms including 24 Ag, and 8 Bi and 24 O atoms, and the corresponding BZ is shown in Fig.~\ref{fig:s4}b.

SG 80 hosts only Weyl phonons at P point (the high-symmetry points are defined in Ref.~\cite{28}) with one two-dimensional irrep R1R2 of the abstract group $G_{2}^1$ in Table~\ref{tab:SEWP}. In $G_{2}^1$, there is only one generator $C_{2z}$. However, by definition it is easy to deduce that the combination operator ${\cal T}C_{4z}$ can keep P invariant, although alone $C_{4z}$ and $\cal T$ are not symmetry operator of P, where $\cal T$ notes the time-reversal symmetry. To prove this, we make use of the representation of $C_{4z}$ acting on reciprocal lattice vectors ( $\bg_1,\bg_2,\bg_3$ ) in this body-centered tetragonal Bravais lattice,
\begin{equation}
\begin{split}
& C_{4z} \left(\begin{array}{c}
\bg_1 \\ \bg_2 \\ \bg_3
\end{array}\right)=
\left[\begin{array}{ccc}
1 & 0 & -1 \\
1 & 0 &  0 \\
1 & -1 & 0 \end{array}\right]
\left(\begin{array}{c}
\bg_1 \\ \bg_2 \\ \bg_3
\end{array}\right)
; \\
\end{split}
\end{equation}
So that at the P point ($\frac{1}{4},\frac{1}{4},\frac{1}{4}$),
\begin{equation}
\begin{split}
&C_{4z} \ {\rm{P}}=(\frac{1}{4},\frac{1}{4},\frac{1}{4})
\left[\begin{array}{ccc}
1 & 0 & -1 \\
1 & 0 &  0 \\
1 & -1 & 0 \end{array}\right] = (\frac{3}{4},-\frac{1}{4},-\frac{1}{4});\\
&{\cal T}(C_{4z}\ {\rm{P}})=(-\frac{3}{4},\frac{1}{4},\frac{1}{4})={\rm P} \text{ (up to integer reciprocal lattice vectors)}
\end{split}
\end{equation}

In this body-centered tetragonal Bravais lattice, $C_{4z}$ acting on the primitive lattice vectors ($\bt_1,\bt_2,\bt_3$) are presented in Ref.~\cite{28}:
\begin{equation}
\begin{split}
& C_{4z} \left(\begin{array}{ccc}
\bt_1 &\bt_2 &\bt_3
\end{array}\right)=
\left(\begin{array}
{ccc}
\bt_1 &\bt_2 &\bt_3
\end{array}\right)
\left[\begin{array}{ccc}
0 & 1 & 0 \\
0 & 1 & -1 \\
-1 & 1 & 0 \end{array}\right],
\end{split}
\end{equation}
Thus,
\begin{eqnarray*}
{\bf 4_1}^2&=& \{C_{4z}|\frac{3}{4}\frac{1}{4}\frac{1}{2}\}^2= \{C_{2z}|1, 0, 0\}\\
{\bf 4_1}^4&=& \{C_{4z}|\frac{3}{4}\frac{1}{4}\frac{1}{2}\}^4= \{E|1, 1, 0\}\\
\end{eqnarray*}
At the P point ($\frac{1}{4},\frac{1}{4},\frac{1}{4}$), the pure translation operator $\{E|1,1,0\}$  is expressed as $e^{2i\pi(1+1+0)/4}=-1$.  In addition, since ${\cal T}^4=1$, we get $({\cal T}{\bf 4_1})^4=-1$, which yields that all the phonon bands have to be doubly degenerate at the P point.
We have checked that there is no symmetry-protected degeneracy on the high-symmetry planes/lines crossing the P point.

The phononic dispersions and DOS of Ag$_3$BiO$_3$ from the \emph{ab initio} calculations are shown in Fig.~\ref{fig:s4}c. It is clearly seen that two-fold SEWPs are localized at the high-symmetry point P.
By fitting the two phonon bands of the lower highlighted WP of P in Fig.~\ref{fig:s4}(c), the $v_1,~v_2$ and $v_3$ coefficients in Eq.~(\ref{eq:H80}) are given 
as 0.12, 0.12 and -1.43 THz$\cdot$\AA.

\clearpage

\begin{figure}[!htb]
\centering
\includegraphics[width=11.0 cm]{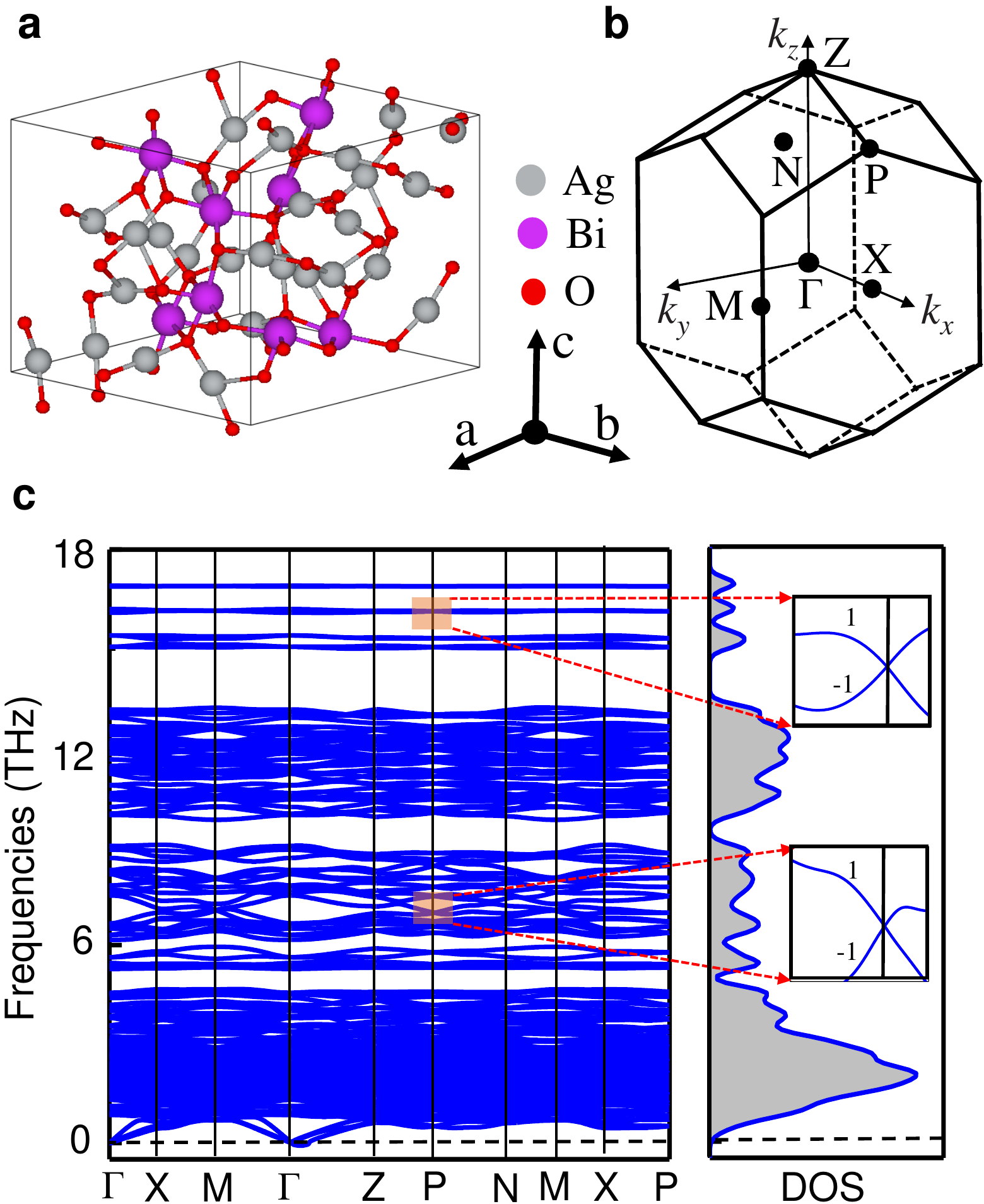}
\caption{A realistic material candidate Ag$_3$BiO$_3$ in SG 80. \textbf{a,} A unit cell contains 56 atoms. \textbf{b,} The first BZ of Ag$_3$BiO$_3$. \textbf{c,} The phonon dispersions of Ag$_3$BiO$_3$ along high-symmetry directions. It is clearly seen that the two-fold Weyl phonons are localized at the high-symmetry point P of the first BZ, and the Chern numbers ({$\pm$}1) of the linear crossing bands are highlighted in the insets of the right panel of the figure \textbf{c}, supporting the appearance of the SEWPs in the material.} \label{fig:s4}
\end{figure}


\subsubsection*{\normalsize{3. The SEWPs in the material LiAuO$_2$ in SG 98}}

The crystallographic data of LiAuO$_2$ are adopted from Ref.~\cite{sf1} and the primitive cell is shown in Fig.~\ref{fig:s5}a, where the green, yellow and red atoms stand for Li, Au and O atoms, respectively. LiAuO$_2$ belongs to the body-centered tetragonal structure with SG I4$_1$22 (No. 98). Each primitive cell contains 8 atoms with 2 Li, and 2 Au and 4 O atoms, and the corresponding BZ is shown in Fig.~\ref{fig:s5}b.

SG 98 hosts only Weyl phonons at the P point (the high-symmetry points are defined in Ref.~\cite{28}), and it hosts one irrep of the AG $G_{16}^7$ in Table~\ref{tab:SEWP}. This SG has a body-centered tetragonal Bravais lattice. The operators $C_{2y}$ and $C_{2z}$ acting on the primitive lattice vectors ($\bt_1,\bt_2,\bt_3$) are presented~\cite{28} as follows,
\begin{equation*}
 \begin{split}
& C_{2z} \left(\begin{array}{ccc}
\bt_1 &\bt_2 &\bt_3
\end{array}\right)=
\left(\begin{array}
{ccc}
\bt_1 &\bt_2 &\bt_3
\end{array}\right)
\left[\begin{array}{ccc}
0 & 1 & -1 \\
1 & 0 & -1 \\
0 & 0 & -1 \end{array}\right];\\
& C_{2y} \left(\begin{array}{ccc}
\bt_1 &\bt_2 &\bt_3
\end{array}\right)=
\left(\begin{array}
{ccc}
\bt_1 &\bt_2 &\bt_3
\end{array}\right)
\left[\begin{array}{ccc}
0 & -1 & 1 \\
0 & -1 & 0 \\
1 & -1 & 0 \end{array}\right],
 \end{split}
\end{equation*}
Thus,
\begin{eqnarray*}
AB&=& \{C_{2z}|000\}\{C_{2y}|0\frac{1}{2}\frac{1}{2}\}= \{C_{2x}|0,-\frac{1}{2},-\frac{1}{2}\}\\
BA&=& \{C_{2y}|0\frac{1}{2}\frac{1}{2}\}\{C_{2z}|000\}=\{C_{2x}|0,\frac{1}{2},\frac{1}{2}\} \\
&=& \{E|0,1,1\} AB
\end{eqnarray*}
At the P point ($\frac{1}{4},\frac{1}{4},\frac{1}{4}$), the pure translation operator $\{E|0,1,1\}$ is expressed as $e^{2i\pi(0+1+1)/4}=-1$. Therefore, we get $\{A,B\}=0$, which yields that all the phonon bands have to be degenerate at the P point.

The phonon dispersions and DOS of LiAuO$_2$ from the \emph{ab initio} calculations are shown in Fig.~\ref{fig:s5}c. The two-fold SEWPs are localized at the high-symmetry point P.

\begin{figure}[!htb]
\centering
\includegraphics[width=11.0 cm]{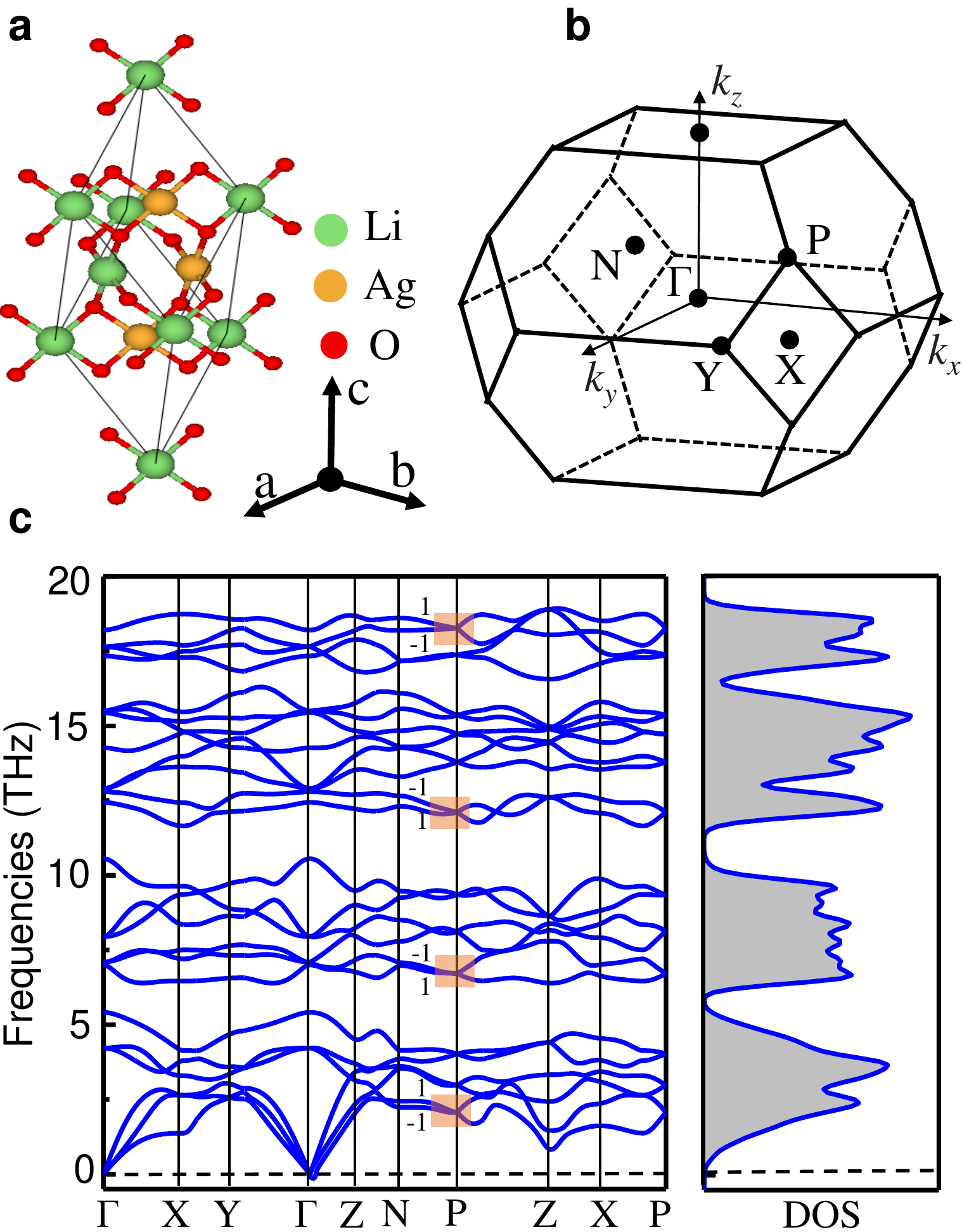}
\caption{A realistic material candidate LiAuO$_2$ in SG 98. \textbf{a,} A unit cell contains 8 atoms. \textbf{b,} The first BZ of LiAuO$_2$. \textbf{c,} The phonon dispersions of LiAuO$_2$ along high-symmetry directions, in which the Chern number ($\pm$1) of some linear crossing bands are highlighted. It is clearly seen that the two-fold SEWPs are localized at the high-symmetry point P of the first BZ.} \label{fig:s5}
\end{figure}
\clearpage

\subsubsection*{\normalsize{4. The SEWPs in the material K$_2$Pb$_2$O$_3$ in SG 199}}

The crystallographic data of K$_2$Pb$_2$O$_3$ are adopted from Ref.~\cite{sf1} and the primitive cell is shown in Fig.~\ref{fig:s6}a, where the purple, black and red atoms stand for K, Pb and O atoms, respectively. K$_2$Pb$_2$O$_3$ belongs to the body-centered cubic structures with SG I2$_1$3 (No. 199)~\cite{sf3}. Each primitive cell contains 6 atoms with 4 K, and 4 Pb and 6 O atoms, and the corresponding BZ is shown in Fig.~\ref{fig:s6}b.

It is noted that K$_2$Pb$_2$O$_3$ belongs to the same SG 199 which has already been studied as an example in the main text, thus the symmetry analysis and the effective $k\cdot p$ model can be referred to the descriptions in the main text. The phonon dispersions and DOS of K$_2$Pb$_2$O$_3$ from the \emph{ab initio} calculations are shown in Fig.~\ref{fig:s6}c. It is clearly seen that two-fold SEWPs and three-fold spin-1 SEWPs are localized at the high-symmetry points P and H, respectively.

\begin{figure}[!htb]
\centering
\includegraphics[width=11.0 cm]{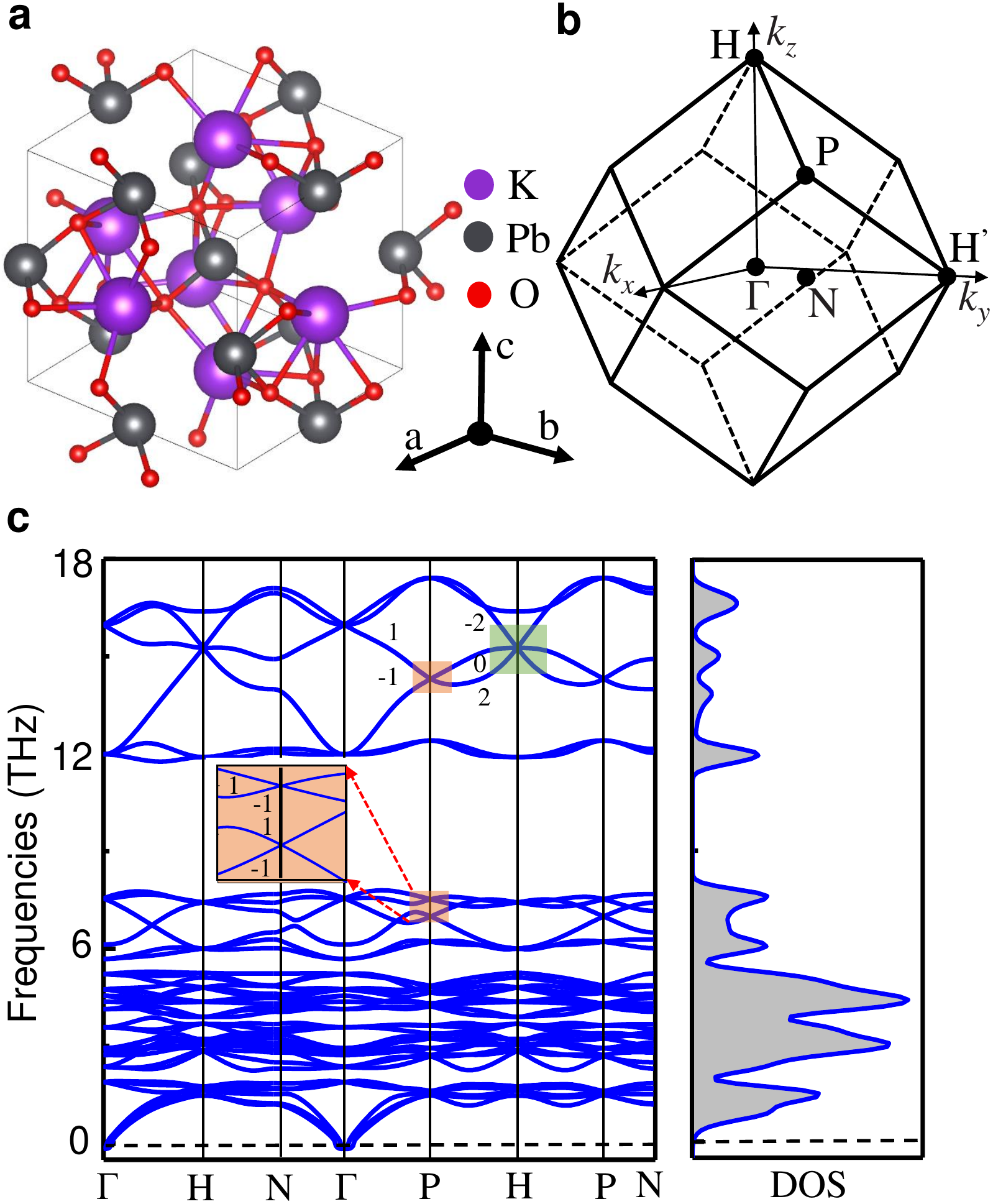}
\caption{A realistic material candidate K$_2$Pb$_2$O$_3$ in SG 199. \textbf{a,} Cubic unit cell contains 6 atoms. \textbf{b,} The first BZ of K$_2$Pb$_2$O$_3$. \textbf{c,} The phonon dispersions of K$_2$Pb$_2$O$_3$ along high-symmetry directions. Similar to K$_2$Sn$_2$O$_3$ studied in the main text, the two-fold Weyl phonons are localized at the high-symmetry point P.} \label{fig:s6}
\end{figure}
\clearpage
\subsubsection*{\normalsize{5. The SEWPs in the material SiO$_2$ in SG 210}}

The crystallographic data of SiO$_2$ are obtained from Ref.~\cite{sf1} and the primitive cell is shown in Fig.~\ref{fig:s7}a, where the blue and red atoms stand for Li and O atoms, respectively. SiO$_2$ belongs to the face-centered cubic structures with SG F4$_1$32 (No. 210). Each primitive cell contains 36 atoms including 12 Si, and 24 O atoms, and the corresponding BZ is shown in Fig.~\ref{fig:s7}b.

SG 210 hosts only Weyl phonons at the W point (the high-symmetry points are defined in Ref.~\cite{28}) with a two-dimensional irrep R10 of the abstract group $G_{16}^7$ in Table~\ref{tab:SEWP}. This space group is a face-centered cubic Bravais lattice. The operators $C_{2x}$ and $ C_{2f}$ acting on the primitive lattice vectors ($\bt_1,\bt_2,\bt_3$) are presented~\cite{28} as below:
\begin{equation}
\begin{split}
& C_{2x} \left(\begin{array}{ccc}
\bt_1 &\bt_2 &\bt_3
\end{array}\right)=
\left(\begin{array}
{ccc}
\bt_1 &\bt_2 &\bt_3
\end{array}\right)
\left[\begin{array}{ccc}
-1 & -1 & -1 \\
0 & 0 & 1 \\
0 & 1 & 0 \end{array}\right];\\
& C_{2f} \left(\begin{array}{ccc}
\bt_1 &\bt_2 &\bt_3
\end{array}\right)=
\left(\begin{array}
{ccc}
\bt_1 &\bt_2 &\bt_3
\end{array}\right)
\left[\begin{array}{ccc}
-1 & 0 & 0 \\
0 & 0 & -1 \\
0 & -1 & 0 \end{array}\right],
\end{split}
\end{equation}
Thus,
\begin{eqnarray*}
AB&=& \{C_{2x}|0 0 0\} \{C_{2f}|\frac{1}{4}\frac{1}{4}\frac{1}{4}\}=\{C_{2d}|-\frac{3}{4} \frac{1}{4}\frac{1}{4}\} \\
BA&=& \{C_{2f}|\frac{1}{4}\frac{1}{4}\frac{1}{4}\}\{C_{2x}|0 0 0\} = \{C_{2d}|\frac{1}{4} \frac{1}{4}\frac{1}{4}\}\\
&=& \{E|1,0,0\} AB
\end{eqnarray*}
where $C_{2d}$ is the Jone's symbol defined in cubic lattice as
\begin{equation}
\begin{split}
& C_{2d} \left(\begin{array}{ccc}
\bt_1 &\bt_2 &\bt_3
\end{array}\right)\equiv
\left(\begin{array}
{ccc}
\bt_1 &\bt_2 &\bt_3
\end{array}\right)
\left[\begin{array}{ccc}
1 & 1 & 1 \\
0 & -1 & 0 \\
0 & 0 & -1 \end{array}\right];
\end{split}
\end{equation}
It's easy to prove that $C_{2x}C_{2y}$=$C_{2y}C_{2x}$=$C_{2d}$.
At the W point ($\frac{1}{2},-\frac{1}{4},\frac{3}{4}$), the pure translation operator $\{E|1,0,0\}$ is expressed as $e^{2i\pi/2}=-1$. Therefore, we get $\{A,B\}=0$, which yields all the phonon bands have to be degenerate at the P point.
We have checked that there is no symmetry-protected degeneracy on the high-symmetry planes/lines that cross the W point. Especially, we have to exclude those containing inversion symmetry and improper rotational symmetries.

The phonon dispersions and DOS of SiO$_2$ from the \emph{ab initio} calculations are shown in Fig.~\ref{fig:s7}c. It is clearly seen that the two-fold SEWPs are located at the high-symmetry point W, which is in agreement with the above results from the symmetry analysis. It is noted that the phononic spectra of SiO$_2$ in SG 210 show many imaginary frequencies, because it isn't a stable crystal.

\begin{figure}[!htb]
\centering
\includegraphics[width=11.0 cm]{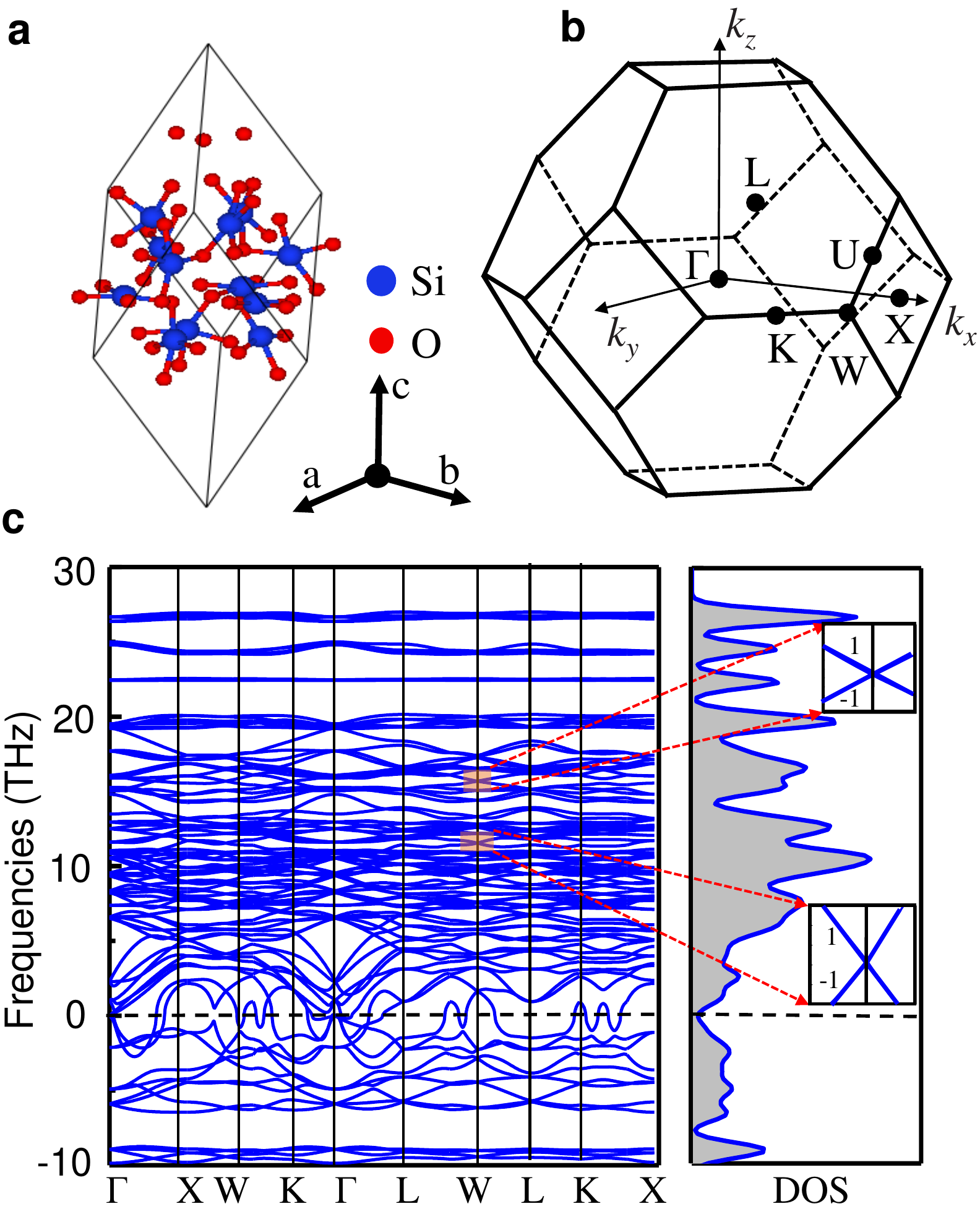}
\caption{A realistic material candidate SiO$_2$ in SG 210. \textbf{a,} A cubic unit cell contains 36 atoms. \textbf{b,} The first BZ of SiO$_2$. \textbf{c,} The phonon dispersions and DOS of SiO$_2$ along high-symmetry directions, and the Chern number ($\pm$1) of some linear crossing bands are highlighted in the right panel. It is clear that the symmetry-enforced two-fold Weyl phonons are localized at the high-symmetry point W of the first BZ.} \label{fig:s7}
\end{figure}
\clearpage

\subsubsection*{\normalsize{6. The SEWPs in the material $K_6$ Carbon in SG 214}}

The primitive cell of $K_6$ Carbon in SG 214 is described in Fig.~\ref{fig:s8}a, where the brown atoms stand for C atoms. $K_6$ Carbon belongs to the body-centered cubic structures with SG I4$_1$32 (No. 214), which are in good agreements with previous calculation results~\cite{sf4}. Each primitive cell contains 6 C atoms, and the corresponding BZ is shown in Fig.~\ref{fig:s8}b.

Similar to SG 199, SG 214 hosts only Weyl phonons at the P point (the high-symmetry points are defined in Ref.~\cite{28}, even though it hosts three different two-dimensional irreps of the abstract group $G_{48}^3$ in Table~\ref{tab:SEWP}. Since the exactly same position and symmetry of the P point in SG 214 and SG 199, we can easily get the same anti-commutation relation $\{A,B\}=0$ by same process where A:$\{{C_{2x}|\frac{1}{2}\frac{1}{2}0}\}$,B:$\{{C_{2z}|\frac{3}{2}1\frac{1}{2}}\}$. The anti-commutation relation yields that all the phonon bands have to be degenerate at the P point.
We have checked that there is no symmetry-protected degeneracy on the high-symmetry planes/lines that cross the P point. Especially, we have to exclude those containing inversion symmetry and improper rotation symmetries.

The phonon dispersions and DOS of $K_6$ carbon from the \emph{ab initio} calculations are shown in Fig.~\ref{fig:s8}c. It is clearly seen that two-fold SEWPs are localized at the high-symmetry point P.
By fitting the two phonon bands of the middle highlighted WP of P in Fig.~\ref{fig:s8}(c), the $v_1,~v_2$ and $v_3$ coefficients in Eq.~(\ref{eq:H199}) are given 
as 1.66, 1.66 and 1.66 THz$\cdot$\AA.

\begin{figure}[!htb]
\centering
\includegraphics[width=11.0 cm]{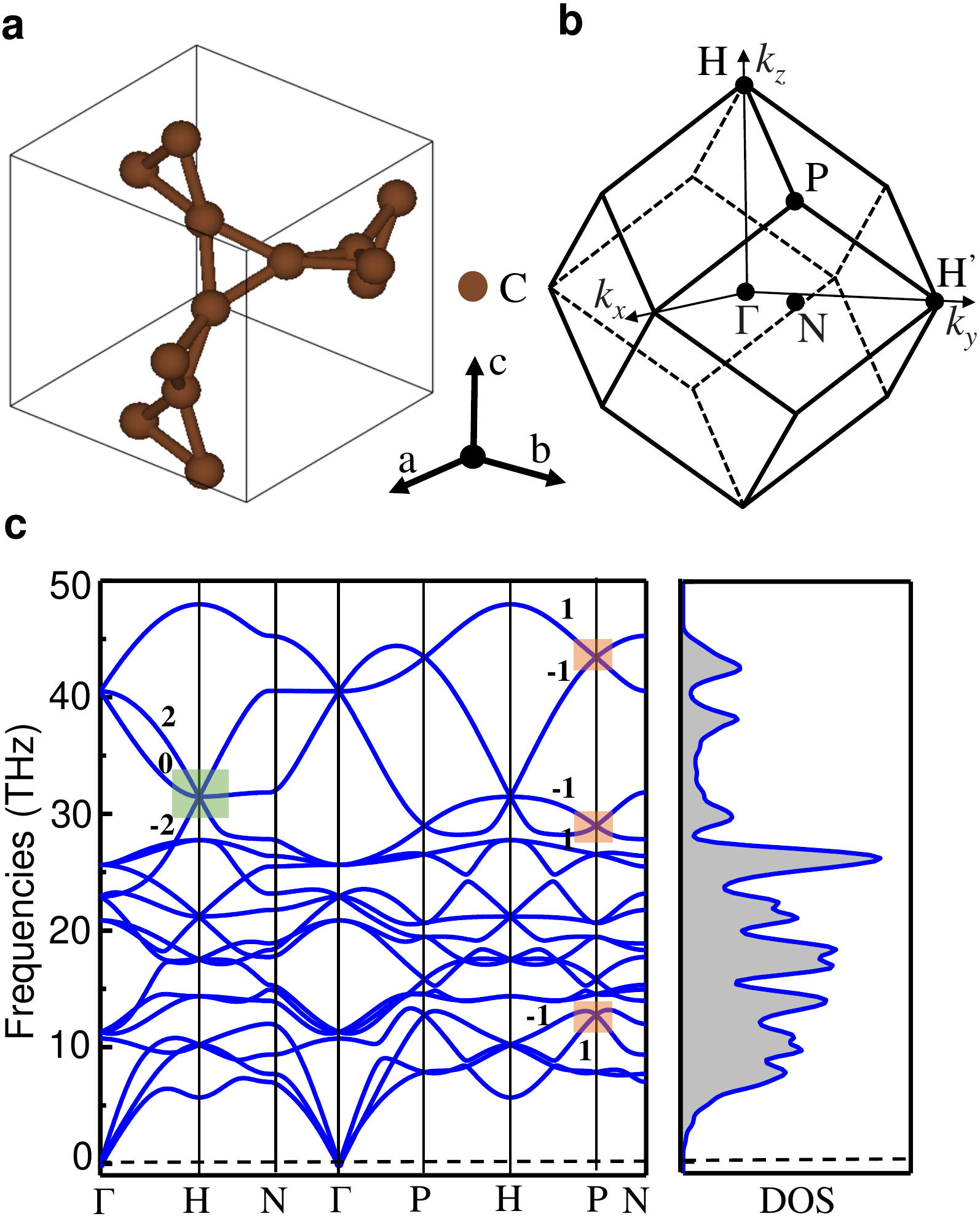}
\caption{A realistic material candidate $K_6$ Carbon in SG 214. \textbf{a,} A cubic unit cell contains 6 C atoms. \textbf{b,} The first BZ of $K_6$  Carbon. \textbf{c,} The phonon dispersions of $K_6$  Carbon along high-symmetry directions. The Chern numbers (0 and $\pm$2) of the three linear crossing bands around the high-symmetry point H, and the Chern numbers ($\pm$1) of the two linear crossing bands at the high-symmetry point P are highlighted.} \label{fig:s8}
\end{figure}
\clearpage

\end{widetext}


\end{document}